\documentclass[12pt]{article}
\usepackage{amsmath,amssymb,amsfonts,graphicx,empheq} 
\usepackage{slashed,tabularx}
\usepackage{young}
\usepackage[all]{xy}

\usepackage[usenames,dvipsnames]{xcolor}
\usepackage[colorlinks=true, pdfstartview=FitV, linkcolor=black, citecolor=black, urlcolor=black]{hyperref}

\def\leigh{Robert G. Leigh}
\def\weiss{Alexander B. Weiss}
\def\onkar{Onkar Parrikar}
\def\uiucaddress{\small Department of Physics, University of Illinois, 1110 W. Green St., 
Urbana IL 61801-3080, U.S.A. }
\def\title{\Large { The Holographic Geometry of the Renormalization Group\\ and Higher Spin Symmetries}}

\parskip=\baselineskip
\renewcommand{\baselinestretch}{1.2}
\newcommand\pasl{{\slash\!\!\!{\partial}}}
\newcommand\cc[1]{#1^{^{\kern-6pt \circ}}\kern2pt}

\newcommand{\pa}{\partial}

\newcommand{\beq}{\begin{equation}}
\newcommand{\eeq}{\end{equation}}
\newcommand{\beqn}{\begin{eqnarray}}
\newcommand{\eeqn}{\end{eqnarray}}

\def\dalemb#1#2{{\vbox{\hrule height .#2pt
\hbox{\vrule width.#2pt height#1pt \kern#1pt
\vrule width.#2pt}
\hrule height.#2pt}}}

\newcommand{\cut}{M}
\newcommand{\rgla}{\cut d_{\cut}}
\newcommand{\bl}{\cdot}
\newcommand{\Conn}{W}
\newcommand{\Scal}{A}
\newcommand{\ScalM}{\Pi}
\newcommand{\ConnM}{\Pi}
\newcommand{\cScal}{\mathcal{A}}
\newcommand{\cO}{\mathcal O}
\newcommand{\cConn}{\mathcal{W}}
\newcommand{\cConnM}{\mathcal{P}}
\newcommand{\cScalM}{\mathcal{P}}
\newcommand{\tpsi}{\tilde{\psi}}
\newcommand{\gmat}{\gamma}
\newcommand{\cL}{\mathcal{L}}
\newcommand{\re}{\mathbb{R}}
\newcommand{\ue}{\underline{e}}
\newcommand{\Cont}{\widehat{W}}
\newcommand{\cCont}{\widehat{\mathcal{W}}}
\newcommand{\pot}{U}
\newcommand{\potfunc}{\mathcal\pot}

\newcommand{\fr}{\mathbb}
\newcommand{\cD}{\mathcal{D}}
\newcommand{\cF}{\mathcal{F}}
\newcommand{\alp}{\Delta}
\newcommand{\uM}{\alpha}
\newcommand{\cG}{\mathcal{G}}
\newcommand{\PConn}{\boldsymbol \omega}
\newcommand{\bd}{\boldsymbol d}

\begin{document}

\begin{center}
\title
\end{center}
\vskip 1 cm
\centerline{
	{\bf
	{\leigh, \onkar\ and \weiss}}
	}
\vspace{.5cm}
\centerline{\it \uiucaddress}

\begin{abstract}
We consider the Wilson-Polchinski exact renormalization group applied to the generating functional of single-trace operators at a free-fixed point in $d=2+1$ dimensions. By exploiting the rich symmetry structure of free field theory, we study the geometric nature of the RG equations and the associated Ward identities. The geometry, as expected, is holographic, with $AdS$ spacetime emerging correspondent with RG fixed points. The field theory construction gives us a particular vector bundle over the $d+1$-dimensional RG mapping space, called a jet bundle, whose structure group arises from the linear orthogonal bi-local transformations of the bare fields in the path integral.
The sources for quadratic operators constitute a connection on this bundle and a section of its endomorphism bundle. 
Recasting the geometry in terms of the corresponding principal bundle, we arrive at a structure remarkably similar to the Vasiliev theory, where the horizontal part of the connection on the principal bundle is Vasiliev's higher spin connection, while the vertical part (the Faddeev-Popov ghost) corresponds to the $S$-field. The Vasiliev equations are then, respectively, the RG equations and the BRST equations, with the RG beta functions encoding bulk interactions. Finally, we remark that a large class of interacting field theories can be studied through integral transforms of our results, and it is natural to organize this in terms of a large $N$ expansion.
\end{abstract}

\pagebreak

\parskip= 2pt
\renewcommand{\baselinestretch}{.2}
\tableofcontents
\parskip=10pt
\renewcommand{\baselinestretch}{1.2}
\section{Introduction}

One of the most appealing aspects of gauge/gravity duality (or holography) is its interpretation as a geometrization of the renormalization group (RG) of quantum field theories. In this picture, scale transformations in the field theory correspond to movement in the extra `radial' direction, and specific RG trajectories correspond to specific geometries, which are asymptotically $AdS$ if the RG flow begins or ends near a fixed point. The precise details of this interpretation are somewhat controversial, and many variants exist in the literature. Early papers \cite{deBoer:1999xf,deBoer:2000cz} on the subject noted the relationship between RG flow and Hamilton-Jacobi theory of the bulk radial evolution. The literature on the subject is vast but some highlights include \cite{Alvarez:1998wr,Akhmedov:1998vf,Schmidhuber:1999rb,Henningson:1998gx,Balasubramanian:1999re,Skenderis:2002wp} and the more recent \cite{Koch:2010cy,Faulkner:2010jy,Heemskerk:2010hk,Lee:2009ij,Gomes:2013qza}.

From the perspective of quantum field theory, considerations of the renormalization group usually begin within the context of perturbation theory, naturally interpreted in terms of deformations away from the free RG fixed point. Indeed, the `exact renormalization group' (ERG) originally formulated by Polchinski  \cite{Polchinski:1983gv} was constructed within the confines of a path integral over bare elementary fields with (regulated) canonical kinetic terms corresponding to the free fixed point. Thus both the power and the curse of ERG is that it is formulated in terms of the free fixed point. One of the hallmarks of holography is that it pertains to a quite opposite limit, in which simple geometric constructions in the bulk correspond to strongly coupled dynamics in the dual field theory. So on the face of it, one might expect very little relationship to exist between the exact renormalization group and holography.

Nonetheless, the geometrization picture begs for such a relationship to exist. In somewhat vague terms, one might expect that passing towards weaker couplings on the field theory side should correspond to some sort of non-geometric version of string theoretic (or M-theoretic) constructions. To be more precise, simple geometric theories in spacetime arise from string theory in a limit in which the string scale $\alpha'$ is small, the mass gap between gravitational fields (and their partners) and other string modes being large. One might then expect that a way to non-geometry in string theory is to take $\alpha'$ large. Unfortunately, very little is known reliably about such a limit. One can say that within the usual spacetime picture, apparently a great many fields of arbitrarily high spin are becoming light (see for example \cite{Sagnotti:2003qa} and \cite{Engquist:2005}). That this can be thought of in effective field theory terms is doubtful.

We do however have one data point: a classical theory involving an infinite number of higher spin gauge fields exists on $AdS$ geometry, a subject primarily developed by Vasiliev (see for example \cite{Vasiliev:1995dn, Vasiliev:2012vf, Vasiliev:1999ba} and the reviews \cite{Bekaert:2005vh, Giombi:2012ms}.). It has been widely speculated that this has something to do with the $\alpha'\to\infty$ limit of string theory. In fact, a conjectured duality between 3d vector models and higher spin theory \cite{Klebanov:2002ja} is well known (see also \cite{Sezgin:2002rt,Leigh:2003gk,Sezgin:2003pt}). Recently, it has been demonstrated \cite{Maldacena:2011jn} that a dual field theory possessing higher spin symmetries must necessarily be free. From these considerations, it seems plausible that holography might be derived from the exact renormalization group, but we should not expect to obtain simple gravitational systems, but rather some sort of higher spin system. Indeed, Douglas et al \cite{Douglas:2010rc} considered this some time ago (see also the followup papers \cite{Zayas:2013qda,Sachs:2013pca} and also \cite{Vasiliev:2012vf,Didenko:2012vh}), suggesting that the higher spin equations of motion ought to be derivable from the exact RG \cite{Polchinski:1983gv} of free field vector-like theories (with global group $G$), the sources for $G$-invariant quadratic operators being related to a (higher spin) connection. In such a picture, there is a connection on some bundle over a $d+1$-dimensional base space, and specific choices of connection should correspond to (higher spin versions of) specific geometries.

Several aspects of this sort of structure must emerge if we are to interpret it as a holographic construction. One of the most basic properties is that $AdS_{d+1}$ should emerge as a geometry associated with an RG fixed point. In geometric language, there must be a specific connection on a bundle over a $d+1$-dimensional topological space, that can be interpreted as being equivalent to having an $AdS$ metric, with its concomitant conformal isometries. But much more challenging is understanding the full diffeomorphism invariance in the bulk $d+1$-space. Any such construction must give rise to this as well.

In this paper, we reconsider and reformulate the scenario of \cite{Douglas:2010rc}, the exact renormalization group (ERG) for field theories whose actions contain arbitrary sources for singlets of a global symmetry. Although we begin with the basic idea of  \cite{Douglas:2010rc}, most of the details of our construction are quite distinct. It turns out that one of the simplest such theories one might consider contains $N$ Majorana fermions in 3 dimensions, with $O(N)$ global symmetry, with an action quadratic in the bare fields. This is the theory that we will study specifically in this paper, although it will be clear that the concepts can be straightforwardly carried over to similar theories in other dimensions, and to scalar field theories as well. As we will explain later in the paper, the full analysis of this theory allows us to construct a large class of interacting theories as well, and we will argue that in the case of the $O(N)$ models, the interacting fixed point is visible at large $N$. 

The free Majorana theory with global symmetry possesses a great many operators in various tensor representations. We choose to ask a specific question of these theories, namely to supply the generating functional of arbitrary `single trace' bi-local operators. It is this question whose answer will be relevant to higher spin theory. We regulate the theory in the same fashion as Polchinski \cite{Polchinski:1983gv} by introducing a cutoff function in the kinetic term. In the context of Majorana fermions, with a single derivative in the kinetic term, this means that we can think in terms of a `regulated derivative operator' and the sources for singlet operators can be organized in such a way that this regulated derivative combines with one of the sources to form a `regulated covariant derivative,' and hence a connection. One of the most important insights that we provide is a precise characterization of the bundle for which this is a connection. {\it What we will find is that the exact renormalization group of the field theory gives rise to a principal bundle over a $d+1$-dimensional space, with the structure group of the bundle corresponding, in the path integral language of the field theory, to bi-local linear transformations of the bare fields.} The RG equations describing the scale dependence of bi-local couplings and correlation functions can be understood as Ward identities associated to these symmetry transformations and map to equations for the curvature of the connection over this bundle. The full field content of the Vasiliev construction is seen to arise in the principal bundle construction, in the sense that the horizontal components of the connection correspond to sources in the field theory, while the vertical components of the connection (the Faddeev-Popov ghosts) correspond to auxiliary pure-gauge degrees of freedom. The mathematical details of the Vasiliev construction can be seen as a specific representation of the structure group. The construction provides a significant geometric interpretation of the pieces of the Vasiliev construction. 

Given that the paper is fairly lengthy and involved, we feel the need to give here a detailed account of the structure and presentation of the paper. In section 2, we formulate the Majorana theory with arbitrary bi-local sources in the classical action. This is structured in such a way that the bi-local sources for the quadratic $O(N)$-singlet operators consist of a Lorentz vector (more precisely, a 1-form) and a pseudoscalar. This structure coincides with the fields appearing in the Vasiliev construction and so is a good starting point. The kinetic term is regulated by a cutoff function, and we refer to the corresponding cutoff derivative operator as $P_F$. We then make the fundamental observation that a change of integration variables in the path integral corresponding to linear, orthogonal, non-local transformations of the bare fields leaves the kinetic term invariant but transforms the vector source as if it were a gauge field, and acts on the pseudoscalar source by conjugation. Since a change of variables in the path integral must be trivial, this leads to a relationship between the generating functional evaluated at different values of the source, i.e., a Ward identity. It is this set of symmetry transformations, which we call $O(L_2)$, for which the vector source is a connection. These symmetry transformations can be extended to include scale transformations as well, the larger group then being called $CO(L_2)$. We note that similar transformations have also been considered previously in \cite{Bekaert:2008sa, Bekaert:2010ky} in the context of higher-spin symmetries. 

In Section 3, we construct the RG equations via a precise sequence of steps involving exact (anomalous) Ward identities and the fundamental property of cutoff independence of the partition function, and show how they may be written as first order differential equations in a $d+1$-dimensional spacetime. These equations form themselves into relations involving the curvature of the connection and the covariant derivative of the pseudoscalar source, with the right hand sides being given by the RG $\beta$-functions. Similarly, the Callan-Symanzik equations for the one-point functions of the singlet operators are derived. A special value of the connection corresponds to the ``pure gauge'' RG flow of the free fixed point, and gives rise  to $AdS_{d+1}$ geometry;  any other connection corresponds to a deformed geometry (including higher spin deformations).

In Section 3.2, we show that the Callan-Symanzik equations are of such a form that they, along with the $\beta$-function equations, admit an interpretation in terms of Hamilton-Jacobi theory, with the radial coordinate of the $d+1$-dimensional space playing the role of `time.' As we mentioned above, this sort of interpretation has been anticipated from the bulk point of view, and it is reassuring that it is a direct consequence of the RG equations of the field theory. The Hamilton-Jacobi theory implies the existence of a corresponding Hamiltonian which is of a special form linear in momenta, and the Hamilton equations derived from it are precisely the full set of RG equations. As well, the appearance of the RG $\beta$-functions in the equations is such that they encode the 3-point functions (in particular) of the field theory, and we show that, at the free fixed point, they are of the expected form. From the bulk point of view, they give rise to the bulk interactions of the higher spin theory.

Section 4 contains a mathematical construction which explains the underlying geometry that emerges from the exact RG equations. (For a previous attempt at understanding the geometry of higher spin theories, see \cite{Sezgin:2011hq}). We introduce and briefly review the concept of {\it jet bundles} to allow us to speak in vector bundle terms. The vector and pseudoscalar sources of the field theory then correspond to a connection on this bundle, and a section of its endomorphism bundle respectively. We then observe that it is useful to think of this connection as being inherited from a connection on the corresponding principal bundle (namely the frame bundle of the jet bundle). The latter connection, of course, also contains a `vertical' piece which in physics language corresponds to the Faddeev-Popov ghosts. These ghost degrees of freedom are pure gauge artifacts that do not have a direct significance in the original field theory, and we conjecture that they should be identified with Vasiliev's auxiliary $S$ field. The equations for $S$ are identified with the BRST equations.  The detailed construction given by Vasiliev involving a $\star$-algebra of $Y$ and $Z$ variables is expected to arise as a particular representation of the structure group of our bundle. 

Thus we arrive at a construction which promises to possess precisely the same content as the Vasiliev theory, although there are a number of differences in the detailed form of the equations, which we highlight. In Section 5, we discuss a number of subsequent issues. First, we organize the bosonic $O(N)$ model in similar terms and note that the most general bi-local sources for singlet quadratic operators consist of a vector and a scalar, again a good starting point for a comparison with the Vasiliev theory. The RG analysis can be worked out along very similar lines, but we do not present the details in this paper. In Section 5.2, we discuss interacting theories. In particular, we note that by taking $N$ large, the partition function of the interacting critical theory can be obtained from our partition function by an integral transform (which constructs a `double trace' deformation by reversing the Hubbard-Stratanovich idea). We conclude the paper with a few additional remarks.


\section{Free Majorana Fermions}
\subsection{Preliminaries}
We consider \(N\) Majorana fermions in $2+1$ dimensional Minkowski spacetime $(\re^{3},\eta)$. We begin with the Dirac action
\beq
S_{Dirac}=\int_x \overline{\psi}^mi\pasl\psi^m
\eeq
As written this has a global $U(N)$ symmetry, where $m,n,...=1,..., N$. Take a basis for $C\ell(2,1)$ as follows
\beq
\gamma^0=i\sigma_2=\epsilon,\ \ \ \  \gamma^1=\sigma_1,\ \ \ \ \gamma^2=\sigma_3
\eeq
where $\sigma_a$ are the $2\times2$ Pauli matrices. This basis is real and the Majorana condition is $\psi^*=\psi$.
We then use the notation $\overline\psi\psi \to \tpsi\psi\equiv\psi_\alpha \varepsilon^{\alpha\beta}\psi_\beta$, etc. Since the Dirac $\psi$ was a fundamental of $U(N)$, the Majorana condition requires that this contract to $O(N)$. The Dirac action then becomes
\beq\label{Majfpaction}
S_{Maj}=\int_x \tpsi^mi\gmat^\mu\pa_\mu\psi^m
\eeq
This action describes the free (Majorana) fermion fixed point. Of course, implicit in the above discussion is the fact that we have picked a frame \(\ue^{(0)}_{a}={\delta_a^{\mu}}\partial_{\mu}\) on $\re^3$, where $a,b\cdots $ are frame indices, and run over the spacetime dimension. We will denote the dual co-frame by $e^a_{(0)}$, and the corresponding metric as 
\beq
g^{(0)}=\eta_{ab}e^a_{(0)}\otimes e^b_{(0)}
\eeq
While we will mostly be interested in $d=3$ Minkowski spacetime, many of our considerations can be generalized straightforwardly to other dimensions, and to non-trivial geometries. For this reason, we will often refer to the spacetime manifold as $M_{d}$, and the background metric as $g^{(0)}$ instead of $\eta$. 

Following Ref. \cite{Polchinski:1983gv}, we regulate the action with a smooth cut-off function $K_F(s)$, which has the property that $K_F(s) \mapsto 1$ for $s<1$ and $K_F(s)\mapsto 0$ for $s>1$. We also wish to add arbitrary bi-local sources for $O(N)$-singlet, single-trace operators\footnote{See Section 5 for comments on interacting theories.}, which in this case are 
\beq
\hat\ScalM(x,y)=\frac12\tpsi^m(x)\psi^m(y),\;\;\;\hat\ConnM^{\mu}(x,y)=\frac12\tpsi^m(x)\gmat^{\mu}\psi^m(y)
\eeq 
The corresponding sources are thus a 0-form \(\Scal(x,y)\) and a 1-form \(\Conn_{\mu}(x,y)\). The resulting action is
\beq\label{MajActFirstForm}
S^{reg.}_{Maj.}=\frac12\int_{x} \tpsi^m(x) K_F^{-1}(-\Box/\cut ^2)i\gmat^{\mu}\pa_\mu\psi^m(x)+\frac12\int_{x,y} \tpsi^m(x)\Big(\Scal(x,y)+\Conn_\mu(x,y)\gmat^{\mu}\Big)\psi^m(y)\nonumber
\eeq
where \(\cut\) is a UV cut-off. Note that because the spinors are Grassmann, the action is only sensitive to the symmetric part of the 0-form $\Scal(x,y)+\Scal(y,x)$ and the antisymmetric part of the 1-form $\Conn_\mu(x,y)-\Conn_\mu(y,x)$. Note also that $\Scal$ is a pseudoscalar. In this form, we see a first indication that the sources $\Scal$ and $\Conn_{\mu}$ are directly related to those found in the Vasiliev higher spin theory. These are just pieces of the full story, as we expect holographically that the sources will combine with  pure gauge modes to form the bulk fields. Nevertheless, the above parametrization seems convenient in order to make contact with higher spin theory. A similar parameterization for the bosonic $O(N)$ model is described in section \ref{sec:bosonic}.\footnote{Note though that in higher dimensions, there are additional single-trace operators, for example $\tpsi^m\gamma^{\mu\nu}\psi^m$, whose sources have no obvious analogue in the Vasiliev higher spin theory. It is for this reason that we specify $d=3$.}

To make further contact with higher spin gauge theory, we note that we may choose to write ``quasi-local" expansions
\beqn
\Scal(x,y) &=& \sum_{s=0}^{\infty} \Scal^{a_1\cdots a_s}(x)\; \pa^{(x)}_{a_1}\cdots \pa^{(x)}_{a_{s}}\delta(x-y)\\
\Conn_{\mu}(x,y) &=& \sum_{s=0}^{\infty} {\Conn_{\mu}}^{a_1\cdots a_{s-1}}(x)\;\pa^{(x)}_{a_1}\cdots \pa^{(x)}_{a_{s-1}}\delta(x-y)
\eeqn
Since the Majorana theory is sensitive to the symmetric part of $\Scal$ and the anti-symmetric part of $\Conn_{\mu}$, we may restrict \(s\) to be even.  From the point of view of the Majorana action, these quasi-local expansions simply mean that we source all \emph{local} single-trace operators, with no prejudice towards the number of derivatives they contain. Nevertheless, we will generally work with arbitrary bi-local sources throughout most of this paper. One of our primary goals is to understand more fully the geometry associated with the bi-local sources, and indeed, in a later section, we will have occasion to re-interpret them in terms of geometric objects on the so-called infinite jet bundles, a construction that allows to think in terms of (infinite dimensional) vector bundles.

Indeed the bi-local nature of the sources leads us to think of them as `matrices' with indices $x,y$, and it is in fact convenient to rewrite the action in the following `matrix' form
\beq\label{MajMatAction}
S^{reg.}_{Maj.}=\int_{x,y}\left[\frac12\tpsi^m(x)\gmat^{\mu} \Big(P_{F;\mu}(x,y)+\Conn_\mu(x,y)\Big)\psi^m(y)+\frac12\tpsi^m(x)\Scal(x,y)\psi^m(y)\right]
\eeq
where we have defined the regulated derivative operator 
\beq
P_{F;\mu}(x,y)= K_F^{-1}(-\Box_{(x)}/\cut ^2)i\pa^{(x)}_\mu\delta(x-y).
\eeq
The introduction of this derivative operator (as opposed to just $\pa^{(x)}_\mu\delta(x-y)$) is ultimately what will tame the non-local character of the theory (we will keep the subscript $F$ throughout the paper to emphasize this, and the reader should regard the $F$ as standing for `cutofF').
Given this matrix form, we will often denote integration simply by a center dot ``$\cdot$'', i.e. 
\beq
(f\bl g)(x,y) = \int_uf(x,u)g(u,y)
\eeq
The corresponding quantum theory is obtained as a path integral
\beq\label{MajPI}
Z[\cut, g^{(0)},\pot, \Scal,\Conn_{\mu}]=(\det \slashed P_F)^{N/2} \int[d\psi] e^{i\pot+iS_{Maj.}^{reg.}[\psi,\Scal,\Conn_{\mu}]}
\eeq
the prefactor included to define the integral, accounting for the fact that $K_F$ cuts off the short-distance modes.\footnote{Since $K_F(s)\to 0$ for $s>1$, the path integral is formally zero due to the integral over $s>1$ field modes unless we include the determinant prefactor, which formally cancels out this effect. The resulting normalization of the path integral will be tracked by introducing a source for the identity operator (i.e., a cosmological constant), which we have denoted by $U$ in eq. (\ref{MajPI}). } Note that we have made explicit the choice of the background metric on spacetime; as has been mentioned before, we are most interested in $g^{(0)}=\eta$, although later we will find it natural to allow its conformal factor to be adjusted. Additionally, we have added in a source $\pot=\int_x\potfunc(x)$ for the identity operator, to keep track of the overall normalization of the path integral. 

We will now show that there is a sense in which $D_\mu(x,y)\equiv P_{F;\mu}(x,y)+\Conn_\mu(x,y)$ should be regarded as a covariant derivative, with $P_F$ playing the role of the ordinary derivative, and $\Conn_\mu$ playing the role of gauge field.

\subsection{The $O(L_2)$ symmetry}
The key observation is that the operator $\hat{\Pi}^{\mu}(x,y)$ is a bilocal current operator, which satisfies a conservation equation. To see this, consider the (connected) vacuum expectation values 
\beq
\Pi^{\mu}(x,y) = -i\frac{\delta}{\delta \Conn_{\mu}(x,y)}\mathrm{ln}\;Z,\;\;\ScalM(x,y) = -i\frac{\delta}{\delta \Scal(x,y)}\mathrm{ln}\;Z \label{PiVEVs}
\eeq
Given the form of the partition function, it is straightforward to show that these satisfy the following conservation equation
\beq
\left[D_{\mu}, \Pi^{\mu}\right]_\cdot + \left[\ScalM,A\right]_\cdot=0. \label{OL2wardidentity1}
\eeq
where $\left[f,g\right]_{\bl}=(f\bl g-g\bl f)$. Inserting (\ref{PiVEVs}) into \eqref{OL2wardidentity1}, multiplying on the left by an infinitesimal antisymmetric parameter \(\epsilon(x,y)\) of compact support, and then taking the functional trace, we obtain 
\beq
\mathrm{Tr}\left\{\left[D_{\mu}, \epsilon\right]\frac{\delta}{\delta\Conn_{\mu}}+\left[\epsilon, A\right]\frac{\delta}{\delta\Scal}\right\}Z[M,g^{(0)},\Scal,\Conn] = 0 \label{OL2Ward1}.
\eeq
The partition function is thus invariant under the transformation
\beq
\delta \Conn_{\mu} = \left[D_{\mu}, \epsilon\right]_\cdot,\;\;\delta A = \left[\epsilon, A\right]_\cdot
\eeq
which resembles a gauge transformation, if we interpret $W_\mu$ as a connection and $A$ as a charged field.
To better elucidate the associated symmetry, we regard (\ref{OL2Ward1}) as a Ward identity, which we now re-derive from a path-integral point of view. To that end, consider a field redefinition 
\beq
\psi^m_\alpha(x)\mapsto \int_y \cL(x,y)\psi^m_\alpha(y).
\eeq
where $\cL: L_2(\re^{d},\eta)\mapsto L_2(\re^{d},\eta)$ is a functional map ($d=3$ in the present case).\footnote{By $L_2(M,g)$ we mean the set of all square integrable functions over the manifold \(M\) with the norm
\[\langle \psi,\psi\rangle_g = \int_M d^dx\sqrt{g(x)}\; \psi(x)\psi(x)\]
} As written, this map acts on the bare fields, the integration variables in the path integral. Formally, the path integral measure in (\ref{MajPI}) is invariant under this linear transformation.\footnote{In terms of the infinitesimal antisymmetric parameter $\epsilon(x,y)$ defined as $\cL(x,y)\simeq\delta(x-y)-\epsilon(x,y)$, this amounts to the assumption that $\epsilon(x,y)$ is trace-class. } For reasons which will become clear soon, we will restrict $\cL$ to  be functionally ``orthogonal," by which we mean
\beq\label{ortho}
(\cL^T\cdot\cL)(x,y)\equiv\int_z \cL(z,x)\cL(z,y)=\delta(x-y).
\eeq
When we need to, we will refer to the group of such orthogonal functional maps\footnote{We can define an orthogonal group $O(V)$ for any vector space $V$ with an inner product, as the group of all endomorphisms on \(V\) which preserves the inner product. This is the source of the notation.} as $O(L_2(\re^d, \eta))$, or simply $O(L_2)$ for short. 
We could obtain a representation in terms of matrices of countable dimension by choosing a suitable discrete basis for \(L_2(\re^d)\). In any case, equation (\ref{ortho}) should be read as 
\beq
``\cL^T\cdot\cL=1".
\eeq 
Let us now consider how the Majorana action behaves under an \(O(L_2)\) transformation
\beqn
S^{reg.}_{Maj.}[\cL\bl\psi, \Scal,\Conn] &=&\frac12 \tpsi^m\bl  \cL^T\bl\gamma^{\mu}(P_{F;\mu}+\Conn_\mu)\bl\cL\bl\psi^m+\frac12\tpsi^m\bl\cL^T\bl\Scal\bl \cL\bl\psi^m\\
&=&\frac12 \tpsi^m \bl\gamma^{\mu} (P_{F;\mu}+\cL^{-1}\bl\Conn_\mu\bl \cL+\cL^{-1}\bl[P_{F;\mu},\cL]_{\bl})\bl\psi^m +\frac12\tpsi^m\bl\cL^{-1}\bl\Scal\bl \cL\bl\psi^m\nonumber\label{OL2transformaction}
\eeqn
where in the last line, we have used the orthogonality condition (\ref{ortho}), allowing us to leave the canonical kinetic operator $P_{F;\mu}$ invariant. Given the assumed invariance of the measure, we arrive at the Ward identity 
\beq
Z\left[M,g^{(0)},\pot,\Scal,\Conn_\mu\right]=Z\left[M,g^{(0)},\pot,\cL^{-1}\bl\Scal\bl \cL,\cL^{-1}\bl\Conn_\mu\bl \cL+\cL^{-1}\bl[P_{F;\mu},\cL]_{\bl}\right] \label{OL2Ward2}
\eeq
So we see that $\Conn_\mu$ behaves like an ``$O(L_2)$ connection'', while $\Scal$ simply conjugates tensorially.
If we now consider the infinitesimal version
 \beq\cL(x,y)\simeq\delta(x-y)-\epsilon(x,y),\eeq the orthogonality condition \eqref{ortho} implies 
 \beq
 \epsilon(x,y)+\epsilon(y,x)=0 \label{OL2condition}
 \eeq
 The infinitesimal version of \eqref{OL2Ward2} is precisely equation \eqref{OL2Ward1}. Note however that we must impose an important constraint on $\epsilon(x,y)$ - since the transformations we are talking about involve mixing elementary field modes, we will require them to have no support (in momentum space) at the cut-off. More precisely, we will impose the condition 
\beq
\left[\epsilon,d_MP_{F;\mu}\right]_{\bl}= 0 \label{constr}
\eeq
where $d_M P_{F;\mu}$ has support only near the cut-off. Physically, the above constraint ensures that we do not mix modes across the UV cutoff. 

Note the significance of equation \eqref{OL2Ward2}. Normally we would say that (\ref{Majfpaction}) is the action of the free fixed point, and that $Z[M,g^{(0)},0,0]$ is the partition function of the regulated theory, with a specific choice of regulated kinetic term. 
The partition function actually depends only on $P_{F;\mu}+W_\mu$ (in particular, the kinetic and source terms have the same tensor structure), and so we could regard the $O(L_2)$ transformation from $W_\mu=0$ to a generic pure gauge connection as a modification of the regulated kinetic term. In other words, \emph{any flat connection (which is gauge equivalent to $\Conn_{\mu}=0$) equally well describes the free fixed point}. It will then be convenient to pull out a flat piece $\Conn^{(0)}_{\mu}$ from $\Conn_{\mu}$:
\beq
\Conn_\mu=\Conn^{(0)}_\mu+\Cont_\mu
\eeq
\beq
dW^{(0)}+W^{(0)}\wedge W^{(0)}=0
\eeq
with $\Cont_\mu$ being a tensor under $O(L_2)$. Here $d=dx^{\mu}\left[P_{F;\mu},\;\;\right]_{\bl}$ is the regulated exterior derivative. For the time-being we will suppress this separation, but it will play a crucial role in the renormalization group analysis.

We have made a choice in splitting $P_{F;\mu}$ and $W_\mu$ apart. Given such a splitting, we would like to consider additional transformations, not contained in $O(L_2)$, which change $P_{F;\mu}$. Indeed, the simplest notion of changing $P_{F;\mu}$ would be to change the cutoff. Such a scale transformation is not contained in $O(L_2)$, and so we will extend that group to a larger one. Of course, changing the cutoff in the regulated kinetic term is precisely the construction of Ref. \cite{Polchinski:1983gv}, and so including that will induce renormalization group transformations. 
Indeed, there is an immediate generalization of (\ref{ortho}) that can be made - instead of considering orthogonal transformations, we can consider transformations orthogonal up to a conformal factor
\beq\label{confortho}
\int_z \cL(z,x)\cL(z,y)=\Omega^2(x)\delta(x-y)
\eeq
We call the group of such transformations $CO(L_2(\re^d))\;$, or $CO(L_2)$ for short. We will mostly be interested in the simpler case, where $\Omega$ is a constant: $\Omega = \lambda^{\Delta_{\psi}}$ with $\Delta_{\psi}=\frac{d-1}{2}$ being the scaling dimension of the bare field \(\psi^m\). The general case is not much harder, and we will comment on it from time to time.

It is convenient at this stage to introduce a conformal factor $z$ in the background metric: $g^{(0)}_{\mu\nu}=z^{-2}\eta_{\mu\nu}$. Furthermore, it is also useful to redefine the 0-form by rescaling it: $A_{old} = z A_{new}$. For simplicity, we will drop the subscript  \emph{new} from here on. With these changes, the Majorana action takes the form
\beq
S^{reg.}_{Maj.}=\frac{1}{2z^{d-1}}\int_x\tpsi^m(x)K_{F}^{-1}(-z^2\Box/M^2)i\gamma^{\mu}\partial_{\mu}\psi^m(x)+\frac{1}{2z^{d-1}}\int_{x,y}\tpsi^m(x)\left(\Scal(x,y)+\gamma^{\mu}\Conn_{\mu}(x,y)\right)\psi^m(y)
\eeq
where by $\Box$ we mean the $\eta$-d'Alembertian.
Note that the cutoff function $K_F$ now depends on the conformal factor $z$, and falls off around the scale $\mu=M/z$. Under a $CO(L_2)$ transformation $\psi\mapsto \cL\bl\psi$, the action transforms as
\beqn
S^{reg.}_{Maj}[\cL\psi]&=&\frac{1}{2z^{d-1}}\tpsi^m\bl \cL^T\bl\gamma^{\mu}(P_{F;\mu}+\Conn_\mu)\bl\cL\bl\psi^m+\frac{1}{2z^{d-1}}\tpsi^m\bl\cL^T\bl\Scal\bl \cL\bl\psi^m \nonumber\\
&=&\frac{1}{2z^{d-1}}\tpsi^m \bl \gamma^{\mu}(\cL^T\bl\cL\bl P_{F;\mu}+\cL^T\bl\Conn_\mu \bl\cL+\cL^T\bl[P_{F;\mu},\cL]_{\bl})\bl\psi^m\nonumber\\
&+&\frac{1}{2z^{d-1}}\tpsi^m\bl\cL^T\bl\Scal\bl \cL\bl\psi^m\nonumber\\
&=&\frac{1}{2(\lambda^{-1}z)^{d-1}}\tpsi^m \bl \gamma^{\mu}( P_{F;\mu}+\cL^{-1}\bl\Conn_\mu \bl\cL+\cL^{-1}\bl[P_{F;\mu},\cL]_{\bl})\bl\psi^m \nonumber\\
&+&\frac{1}{2(\lambda^{-1}z)^{d-1}}\tpsi^m\bl\cL^{-1}\bl\Scal\bl \cL\bl\psi^m
\eeqn
Therefore, we find that the action of $CO(L_2)$ can be thought of as an appropriate ``gauge'' transformation on the sources, plus a Weyl transformation of the background metric $z\mapsto \lambda^{-1}z$ (or equivalently $g^{(0)}\mapsto \lambda^2g^{(0)}$) and a rescaling of the cutoff $M \mapsto \lambda^{-1}M$ (note in particular that we have just done a transformation of the bare fields and the argument of $P_F$ has not changed). In addition, we allow for a possible anomaly from the non-invariance of the measure of the path integral, which we will indicate by replacing $\pot \mapsto \widehat{\pot}$. Thus, we arrive at the Ward identity\footnote{At this point, we change our notation slightly, $Z[M,g^{(0)},U,\Scal,\Conn]\mapsto Z[M,z,U,\Scal,\Conn]$, in order to explicitly keep track of the conformal factor $z$. Also note, that although we have allowed the metric of the field theory to change (i.e. by a Weyl transformation), one may equally well think of this as a scale transformation in the sense of a conformal isometry.} 
\beq\label{COL2WardId}
Z\left[\cut,z,\pot,\Scal,\Conn_\mu\right]=Z\left[\lambda^{-1}\cut,\lambda^{-1}z,\widehat{\pot},\cL^{-1}\bl\Scal\bl\cL,\cL^{-1}\bl\Conn_\mu \bl\cL+\cL^{-1}\bl[P_{F;\mu},\cL]\right]
\eeq
Thus, in this sense, the partition function is invariant under this larger $CO(L_2)$ symmetry. The 1-form $\Conn_{\mu}$ transforms like a ``$CO(L_2)$ connection'', while the 0-form $\Scal$ transforms tensorially. Once again if we take $\cL$  infinitesimal, $\cL \simeq \mathbf{1}-\epsilon$ and $\lambda \simeq 1-\varepsilon$ with $\frac{\epsilon+\epsilon^T}{2}=\varepsilon \Delta_{\psi} \mathbf{1}$ (so as to satisfy the orthogonality constraint \eqref{confortho}), then we get
\beq
\delta \Conn_{\mu} = \left[D_{\mu},\epsilon\right]_{\bl},\;\;\delta \Scal = \left[\epsilon,\Scal\right]_{\bl}
\eeq
As in equation \eqref{constr}, we must once again impose 
\beq
[\epsilon,Md_MP_{F;\mu}]=-[\epsilon,zd_zP_{F;\mu}]=0 \label{constr2}
\eeq 
to avoid mixing modes across the cutoff.  

The identity \eqref{COL2WardId} can be extended to the case of \(\lambda\) being a function (rather than a constant). In so doing, one should allow the cutoff to vary in spacetime as well, and introduce a cutoff function appropriately. One possible definition of such a cut-off function is
\beq
K_F\left(-\Box_{x}/M^2\right) \mapsto K_F\left(-\frac{1}{\cut^d\sqrt{g_{(0)}}}\partial_{\mu}(\cut^{d-2}\sqrt{g_{(0)}}\;g_{(0)}^{\mu\nu}\partial_{\nu})\right).
\eeq
This has the feature that a local scale transformation of the metric can be absorbed by a local change in the cutoff. Given this, the partition function would satisfy eq. (\ref{COL2WardId}) locally.

Having described the $O(L_2)$ and $CO(L_2)$ symmetries in some detail, we now move on to study the renormalization group flow out of the free fixed point in light of these symmetries. 

\section{The Renormalization group and Holography}\label{section:RG}

The general principle of Wilsonian renormalization is that the action of a quantum field theory should be thought of as a function of the energy scale at which it is probed. In simple terms, this amounts to having cutoff dependent sources (or couplings) -- this is because, in say lowering the cutoff from $M$ to $\lambda M$ ($\lambda <1$), one is really integrating over the fast modes in the path integral, which consequently changes the values of the couplings, thus making them cutoff dependent. The remarkable feature of the Wilson-Polchinski exact renormalization group \cite{Polchinski:1983gv} is the description of renormalization of a QFT action in terms of a diffusion-like equation, with the cutoff $M$ being the flow parameter.

Alternatively, it is also possible to think of the conformal scale $z$ of the background metric $g^{(0)}=z^{-2}\eta$ as parameterizing the RG flow. In this version, one lowers the cutoff $M\mapsto \lambda M$ by integrating out fast modes, but then performs a scale transformation $g^{(0)}\mapsto \lambda^2g^{(0)}$ (or equivalently $z\mapsto \lambda^{-1}z$) to take the cutoff back to $M$. Naturally, in this case, the conformal factor $z$ acts as the flow parameter, and the sources may be thought of as $z$-dependent. From a geometric point of view, this version of RG is more appealing, and we will adopt it in our discussions below. In the notation introduced in the previous section, we will then regard the sources, $\Scal(z;x,y)$ and $\Conn_{\mu}(z;x,y)$, as functions of $z$. The plan is then to investigate the change in the sources under $z\mapsto \lambda^{-1}z$, while paying special attention to the $CO(L_2)$ symmetry. Following Polchinski \cite{Polchinski:1983gv}, we will be able to write fully-covariant exact differential RG equations by expanding $\lambda$ close to unity.

\begin{figure}[!h]
	\centering
	\includegraphics[width=15cm]{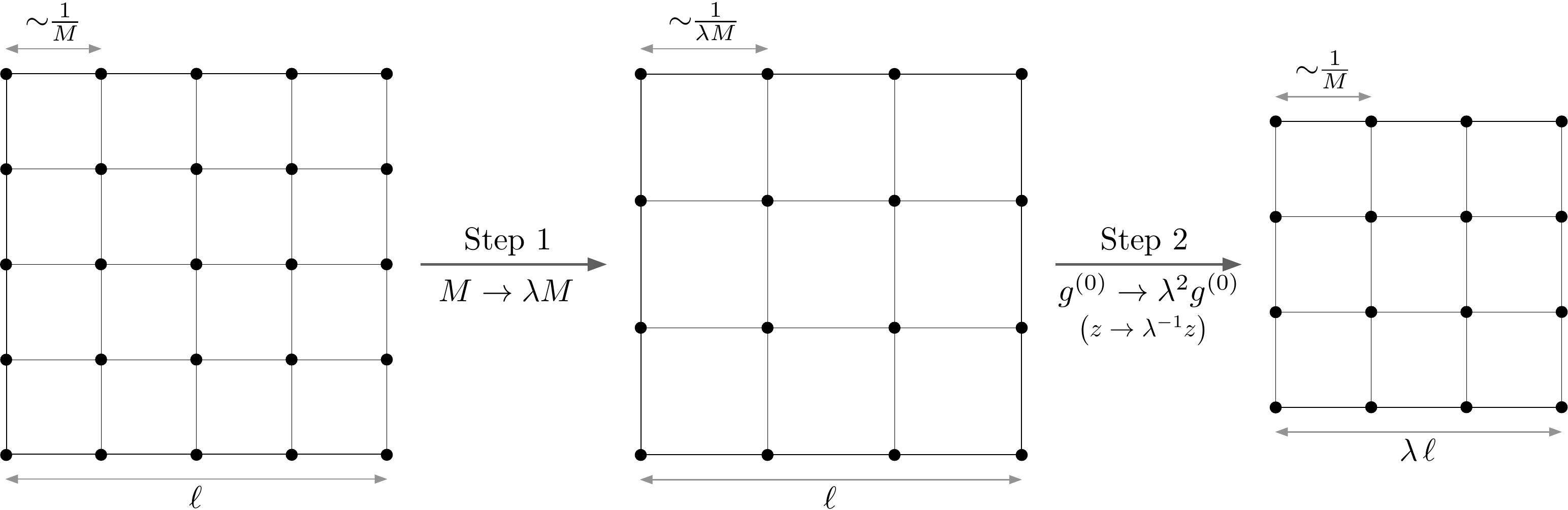}
	\caption{\small{A schematic description of the two step RG process. We have indicated the cutoff in terms of the lattice spacing.}}
	\label{fig:twostepRG}
\end{figure}

\begin{figure}[!h]
	\centering
	\includegraphics[width=6cm]{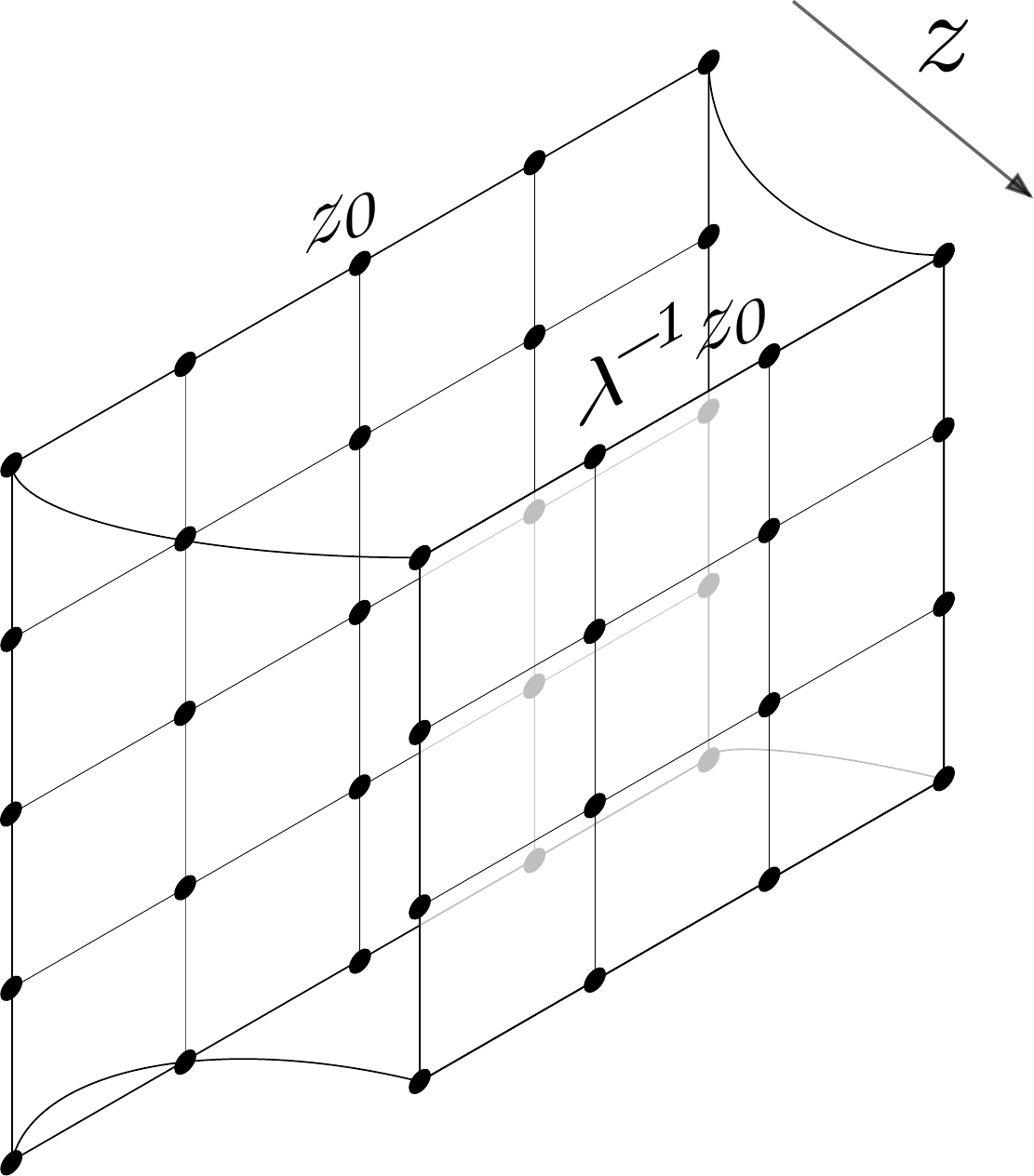}
	\caption{\small{It is useful to think of different values of $z$ as corresponding to different copies of spacetime. From this point of view, a holographic interpretation naturally emerges out of the renormalization group.}}
	\label{fig:holoRG}
\end{figure}

For clarity, we restate the above program as a 2-step process: 

\begin{quote}
\textbf{Step 1}. Lower the cut-off $\cut \to \lambda \cut$, for $\lambda=1-\varepsilon$. This will change the sources, and we label the new sources by $\widetilde\Conn_{\mu}(z)$ and $\widetilde\Scal(z)$.
\beq\label{RGWard1}
	Z[\cut, z,\pot(z),\Scal(z;x,y),\Conn_{\mu}(z;x,y)]
	= Z[\lambda\cut,z,\widetilde{\pot}(z),\widetilde\Scal(z;x,y),\widetilde\Conn_{\mu}(z;x,y)]
\eeq
This result may be worked out in detail using the method of Ref. \cite{Polchinski:1983gv}. The result is discussed in the next section and further details of the calculation may be found in the Appendix.
 
\textbf{Step 2}.  Perform a scale transformation, to bring the cut-off back to $\cut$ while changing the background metric to $g^{(0)}\to \lambda^2 g^{(0)}$, and thus changing the conformal factor $z \to \lambda^{-1}z$.  In the present context, we interpret the scale transformation as a $CO(L_2)$ transformation $\cL$ (with $\cL^T.\cL= \lambda^{2\Delta_{\psi}}\mathbf{1}$).  In addition to this scale transformation, we also have the freedom to translate the spatial coordinates: $x^\mu \rightarrow x^\mu + \varepsilon\xi^\mu$, $y^\mu \rightarrow y^\mu + \varepsilon\xi^\mu$.\footnote{Or, more generally, any isometry of the background metric $g^{(0)}$, so Lorentz diffeomorphisms could also be considered. We choose translations in particular, because they preserve our choice of the background frame, which the fermions couple to.}  
Such a transformation is natural if we regard different values of $z$ as corresponding to different copies of spacetime -- the map between coordinates on one copy to those on another need not be trivial.
\end{quote}
Having performed these two transformations, we now re-label the final sources as $\Conn_{\mu}(\lambda^{-1}z;x+\varepsilon\xi,y+\varepsilon\xi)$ and $\Scal(\lambda^{-1}z;x+\varepsilon\xi,y+\varepsilon\xi)$, and obtain the following equality of partition functions at the same cut-off, but different $z$:
\begin{align}
	&Z[\cut, z,\pot(z),\Scal(z;x,y),\Conn_{\mu}(z;x,y)] \nonumber\\
	&\overset{(1)}{=} Z\left[\lambda\cut,z,\widetilde\pot(z),\widetilde\Scal(z;x,y),\widetilde\Conn_\mu(z;x,y)\right]\nonumber\\
	&\overset{(2)}{=} Z\Big[ \cut , \lambda^{-1}z , \widehat{\widetilde{\pot}}(z) , \cL^{-1}\bl\widetilde\Scal(z;x,y)\bl\cL , 
	                     \cL^{-1}\bl\widetilde\Conn_\mu(z;x,y) \bl\cL+\cL^{-1}\bl[P_{F;	\mu} , \cL ]\Big]\nonumber \label{step2pt2}\\
	&= Z \Big[ \cut, \lambda^{-1}z , \pot(\lambda^{-1}z) , 
	                     \Scal(\lambda^{-1}z;x+\varepsilon\xi , y+\varepsilon\xi) , \Conn_{\mu}(\lambda^{-1}z;x+\varepsilon\xi , y+\varepsilon\xi) \Big]
\end{align}
The first equality is just step one of RG (\ref{RGWard1}), written again for clarity.  The second equality is step two of RG (the $CO(L_2)$ transformation).  This equality includes the notation $\widehat{\widetilde{\pot}}$, denoting the possibility of a $CO(L_2)$ Weyl anomaly, as was mentioned above equation \eqref{COL2WardId}.  The third equality is simply a re-labeling of the source arguments, as described above. The above procedure is indicated pictorially in Figures \ref{fig:twostepRG} and \ref{fig:holoRG}. We note in passing, that since the $CO(L_2)$ transformation can be made local (here, we mean that $\lambda$ can vary in spacetime), the above relations may be regarded as being valid locally, although we will not need to do so. 

Note that we now have a copy of $A$ and $W_\mu$ at each value of $z$. 
Given the interpretation $W_\mu$ as a $CO(L_2)$-connection on spacetime, it is useful to parameterize this connection (now at each $z$) as
\beq
\Conn_{\mu}(z) = \Conn_{\mu}^{(0)}(z)+\Cont_{\mu}(z).
\eeq
where $\Conn^{(0)}(z)=\Conn_{\mu}^{(0)}(z)dx^{\mu}$ is a flat connection
\beq\label{flat}
dW^{(0)}+W^{(0)}\wedge W^{(0)}=0
\eeq
with $d \equiv dx^{\mu}\left[P_{F;\mu},\;\;\right]$. Recall from the previous section, that the reason for separating out the flat piece $\Conn^{(0)}$ from $\Conn_{\mu}$ is that the configuration  $(\Conn_{\mu},A)=(\Conn^{(0)}_{\mu},0)$ is gauge equivalent to the unperturbed free fixed point, and consequently $\Cont_{\mu}$ and $\Scal$ are {\em tensorial} sources for single-trace deformations away from the fixed point. Our primary task is now to describe how under RG, $W_\mu$ naturally evolves into a connection 1-form on a one-higher-dimensional spacetime, namely the \emph{mapping space} of RG, with the extra dimension parametrized by $z$. In fact, as we will see below, $W^{(0)}$ also evolves, in particular, into the $AdS_{d+1}$ connection.

\subsection{Infinitesimal version: RG and Callan-Symanzik equations}

\begin{figure}[!h]
	\centering
	\includegraphics[width=10cm]{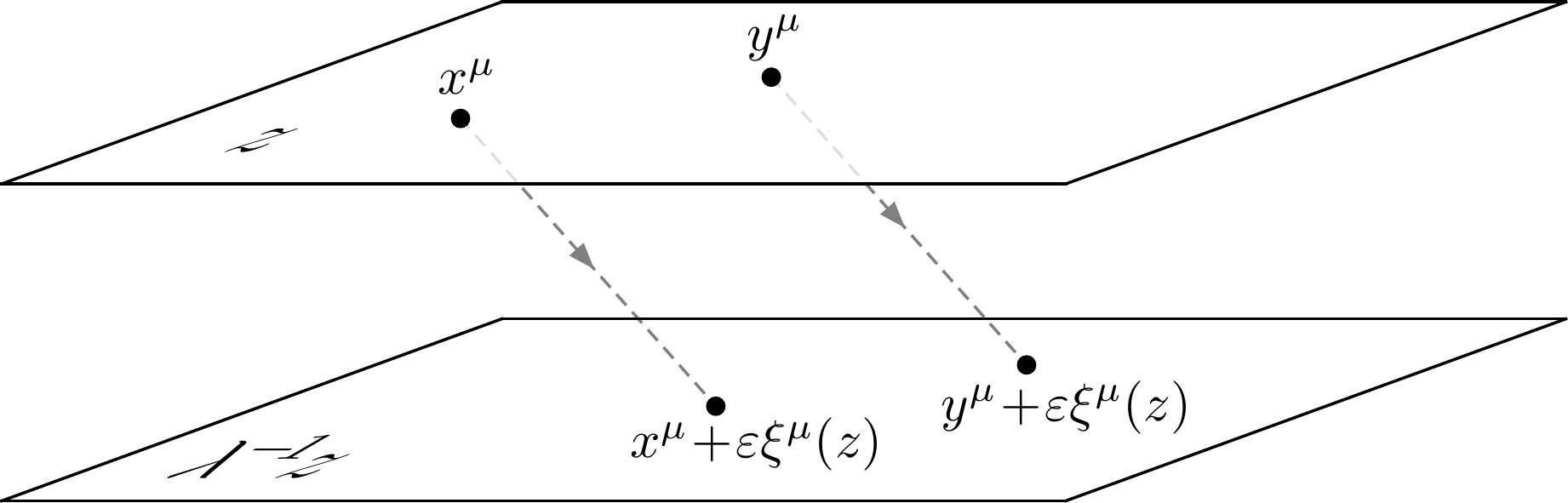}
	\caption{\small{A pictorial representation of the Wilson line interpretation -- the $CO(L_2)$ transformation in step 2 of RG may be thought of as an infinitesimal Wilson line, covariantly transporting sources from $z$ to $z+z\varepsilon$.}}
	\label{fig:translationRG}
\end{figure}

Let us now explore the above relations satisfied by the partition function for infinitesimal transformations: we write $\lambda= 1-\varepsilon$, and \emph{parameterize} the infinitesimal $CO(L_2)$ transformation plus spatial translation appearing in (\ref{step2pt2}) as 
\beq
\cL = \mathbf{1} + \varepsilon z\Conn_z + \varepsilon\xi^\mu W_\mu
\eeq
with $z(\Conn_z+\Conn_z^T)=-2\Delta_{\psi} \mathbf{1}$. Note that we have suggestively re-labeled $\cL$ to indicate that the $\varepsilon$ piece of it should be thought of as containing the $z$-\emph{component} $W_z$ of the connection, while the $\xi$ piece ensures covariance along the transverse directions. Indeed, in this notation, $\cL$ resembles an infinitesimal Wilson-line
\beq
	\cL = \mathbf{1} + \int_0^1dt\;\left(\frac{dz}{dt} W_z+\frac{dx^{\mu}}{dt}W_{\mu}\right)+ O(\varepsilon^2)
\eeq
which covariantly transports sources from $(z;x^{\mu},y^{\mu})$ to $\left(z+z\varepsilon; x^{\mu}+\varepsilon\xi^{\mu},y^{\mu}+\varepsilon\xi^{\mu}\right)$, along the path $(z(t);x^{\mu}(t),y^{\mu}(t))=\left(z+t\varepsilon z,x^{\mu}+t\varepsilon\xi^{\mu},y^{\mu}+t\varepsilon\xi^{\mu}\right)$ (see Figure \ref{fig:translationRG}). $W_z$ is thus a convenient book-keeping device which keeps track of the gauge transformations along the RG flow.
\\ 

Following the two-step RG process outlined above in the infinitesimal case, we get
\beq \label{AMajRG0}
	\Scal(z+\varepsilon z;x+\varepsilon\xi,y+\varepsilon\xi)
	  = \Scal(z;x,y) + \left[ A, \varepsilon z \Conn_z + \varepsilon\xi^\mu W_\mu \right]_\bl + \varepsilon z\beta^{(\Scal)} + O(\varepsilon^2)
\eeq
\beq \label{WMajRG0}
	\Conn_\mu(z+\varepsilon z;x+\varepsilon\xi,y+\varepsilon\xi)
	  = \Conn_\mu (z;x,y) + \left[ P_{F;\mu} + \Conn_\mu , \varepsilon z \Conn_z + \varepsilon\xi^\nu \Conn_\nu \right]_\bl + \varepsilon z \beta^{(\Conn)}_{\mu}
	      + O(\varepsilon^2)
\eeq
where the tensorial RG beta functions are given by\footnote{This is to say that the beta functions as defined transform tensorially under $CO(L_2)$.} 
\beq\label{beta1}
{\beta}^{(\Scal)}(\Scal,\Cont_{\mu})=\Scal\bl \alp^{\mu}\bl \Cont_{\mu}+\Cont_{\mu}\bl \alp^{\mu}\bl \Scal+\varepsilon^{\mu\nu\lambda}\Cont_{\mu}\bl \alp_{\nu}\bl \Cont_{\lambda} 
\eeq
\beq\label{beta2}
{\beta}^{(\Conn)}_{\mu}(\Scal,\Cont_{\mu})= \Scal \bl \alp_{\mu}\bl \Scal+\varepsilon_{\mu\nu\lambda}\left(\Scal \bl \alp^{\nu}\bl \Cont^{\lambda}+\Cont^{\nu}\bl \alp^{\lambda}\bl \Scal\right)+ \Cont_{\nu}\bl \alp^{\nu}\bl \Cont_{\mu}-\Cont_{\nu}\bl \alp_{\mu}\bl \Cont^{\nu}+\Cont_{\mu}\bl \alp_{\nu}\bl \Cont^{\nu}
\eeq
with 
\beq\label{defDelta}
\gamma^{\mu}\alp_{\mu}(x,y) \equiv \cut d_{\cut}\left\{\left(i\slashed{P}_{F}+\slashed{\Conn}^{(0)}\right)^{-1}\right\}(x,y).
\eeq
These are obtained by an explicit computation following \cite{Polchinski:1983gv}, the details of which can be found in Appendix \ref{RGappendix}. We will have more to say about the structure of these beta functions in section \ref{sec:hamiltonjacobi}.

Note that in eq. (\ref{WMajRG0}), the full connection $W$ appears. We may separate this equation into two pieces by requiring that $W^{(0)}(z)$ remains flat along the RG flow at $(1+\varepsilon)z$. In other words, the RG flow of $\Conn_{\mu}^{(0)}$ is pure gauge, and can be expressed in terms of a $CO(L_2)$ transformation parametrized by $W_z^{(0)}$,
\beq
W_{\mu}^{(0)}(z+z\varepsilon;x+\varepsilon\xi,y+\varepsilon\xi)=W_{\mu}^{(0)}(z;x,y)+\left[P_{F;\mu}+W_{\mu}^{(0)},\varepsilon z W^{(0)}_z+\varepsilon\xi^{\nu}W_{\nu}^{(0)}\right]+O(\varepsilon^2) \label{background}
\eeq
Physically, this expresses the fact that the theory is RG invariant at the fixed point. Given equation \eqref{background}, the flow equation for $\widehat{\Conn}_\mu$ can straightforwardly be extracted from (\ref{WMajRG0}).

By continuing the RG process in this way, we may in principle extend $\Conn_{\mu}$ and $\Scal$ from a given value of $z$ to any other value of $z$. As we have seen above, in the process of doing so, the connection 1-form naturally ``grows a leg'' $\Conn_z$ in the $dz$ direction, which keeps track of the gauge transformations along RG flow. It is important to note that given the constraint \eqref{constr2}, the connections $W_{\mu}(z)$ and $W_{\mu}^{(0)}(z)$ as defined by equations \eqref{WMajRG0} and \eqref{background} respectively, transform appropriately even under a $z$-\emph{dependent} $CO(L_2)$ transformation $\tilde{\cL}(z)=\mathbf{1}-\alpha(z)$,
\beq
\delta W_{\mu} =\left[D_{\mu},\alpha\right],\;\;\;\delta W^{(0)}_{\mu} = \left[D^{(0)}_{\mu},\alpha\right]
\eeq
provided we require that $W_z$ and $W_z^{(0)}$ also transform as 
\beq
\delta W_z =\left[D_z,\alpha\right],\;\;\;\delta W^{(0)}_z = \left[D^{(0)}_z,\alpha\right]
\eeq
where we have defined $D_z=d_z+W_z$ and $D_z^{(0)}=d_z+W_z^{(0)}$. Therefore, we now find ourselves in a position to re-interpret $W(z)$ and $W^{(0)}(z)$, as connections over the one-higher-dimensional spacetime $M_{d+1}=\re_+\times \re^{d}$. We will denote these resulting connections over $M_{d+1}$ as ${\cConn}$ and $\cConn^{(0)}$ (to emphasize that they live in the ``bulk''). Further details about the structure of the bundle over which these are connections can be found in Section \ref{sec: jetbundles}. Similarly, the pseudoscalar $A$ extends to a bulk field, which we will denote by ${\cScal}$. 

By comparing the $\varepsilon$ terms on both sides of the RG equations \eqref{background}, \eqref{AMajRG0} and \eqref{WMajRG0} and taking $\varepsilon\to 0$, we obtain
\beqn
\cF^{(0)} &\equiv& \bd\cConn^{(0)}+\cConn^{(0)}\wedge \cConn^{(0)}=0
\\
\label{MajRG1}
i_{\ue^{(0)}_z}\cD\cScal &\equiv & i_{\ue^{(0)}_z}\left(\bd\cScal+\left[\cConn,\cScal\right]\right)=\beta^{(\cScal)}
\\
\label{MajRG2}
i_{\ue^{(0)}_z}\cF & \equiv & i_{\ue^{(0)}_z}\left(\bd \cConn+\cConn\wedge \cConn\right)=\beta_{a}^{(\cConn)}e^a_{(0)}
\eeqn
where we have defined the bulk forms 
$\cConn^{(0)} \equiv \cConn^{(0)}_Idx^I = \cConn^{(0)}_{\mu}dx^{\mu}+\cConn^{(0)}_zdz$, and $\cConn \equiv \cConn_Idx^I = \cConn_{\mu}dx^{\mu}+\cConn_zdz$. In these expressions, we have also introduced the $z$-component $\underline{e}^{(0)}_z\equiv \pa_z+z^{-1}\xi^\mu \partial_{\mu}$ of the boundary frame $\underline{e}^{(0)}_a=\delta_{a}^{\mu}\partial_{\mu}$, the notation $i_{\underline{v}}$ for the interior product of a differential form with the vector field $\underline{v}$, and the {\it regulated bulk exterior derivative} 
\beq 
\bd = dx^{\mu}\left[P_{F;\mu}, \;\;\right]_{\bl}+dz\partial_z.
\eeq 
The corresponding coframe is defined as\footnote{Note that this coframe is merely a choice of basis for 1-forms in the bulk, and should not be confused as having anything to do with $\cConn^{(0)}$. The translation $\xi^\mu$ of the spacetime coordinates as we move in $z$ appears merely as a shift vector in this basis.}
\beq
e^a_{(0)}=\delta^a_{\mu}dx^{\mu}-\xi^a\frac{dz}{z},\;\;\;\;e^z_{(0)}=dz
\eeq
We propose that the RG equations (\ref{MajRG1},\ref{MajRG2}) should be interpreted as the $z$-components of covariant equations
\newcommand{\Beta}{\boldsymbol{\beta}}
\begin{empheq}[box=\fbox]{align}
\cF^{(0)} &= \bd\cConn^{(0)}+\cConn^{(0)}\wedge \cConn^{(0)}=0
\\
\label{MajRG3}
\cD\cScal & = \bd\cScal+\left[\cConn,\cScal\right]=\Beta^{(\cScal)}\\
\label{MajRG4}
\cF &= \bd\cConn+\cConn\wedge \cConn=\Beta^{(\cConn)}
\end{empheq}
where $\beta^{(\cScal)}$ has been promoted to the 1-form $\Beta^{(\cScal)}=\beta^{(\cScal)}e_{(0)}^z+\beta^{(\cScal)}_ae^a_{(0)}$ and similarly $\beta^{(\cConn)}_\mu$ to a 2-form $\Beta^{(\cConn)}=\beta_a^{(\cConn)} e^z_{(0)}\wedge e^a_{(0)}+\beta_{ab}^{(\cConn)} e^a_{(0)}\wedge e^b_{(0)}$. The transverse components of $\Beta^{(\cScal)},\Beta^{(\cConn)}$ not appearing in the original RG equations (\ref{MajRG1},\ref{MajRG2}) are constrained by consistency to satisfy their own flow equations, namely the Bianchi identities
\beq
\cD\Beta^{(\cScal)}=\left[\Beta^{(\cConn)},\cScal\right],\quad\quad
\cD\Beta^{(\cConn)}=0.
\eeq 

Thus, we find that the renormalization group equations organize themselves in terms of covariant equations expressing curvatures in terms of beta functions, with the zeroes of the beta functions corresponding to flat connections. In fact, the first equation simply states that ${\cal W}^{(0)}$, which encodes the pure-gauge RG flow of the free-fixed point, is a flat connection on $M_{d+1}$. At this point, we see the emergence of the $AdS_{d+1}$ spacetime, because in suitable local coordinates, a natural choice for $\cConn^{(0)}$ is given by
\beq
\cConn^{(0)}=-\frac{dz}{z}D(x,y)+\frac{dx^\mu}{z}P_\mu(x,y)\label{AdSconnection}
\eeq
where $P_{\mu}(x,y)=\partial_{\mu}^{(x)}\delta(x-y)$ and $D(x,y) = (x^{\mu}\partial_{\mu}^{(x)}+\Delta_{\psi}+\frac{d}{2})\delta(x-y)$. Note that this choice of $\cConn^{(0)}$ may be regarded as a Cartan connection on $M_{d+1}$, or equivalently as the Maurer-Cartan form of $O(2,d)$, and precisely corresponds to the $AdS_{d+1}$ metric in the Poincar\'e patch (see \cite{Wise:2006sm} for more details). In this way, it seems the renormalization group gives rise to a holographic description.

It is also possible to derive similarly the Callan-Symanzik equations for $\ScalM$ and $\ConnM^{\mu}$ following the 2-step procedure outlined above (see Appendix \ref{RGappendix} for details)
\beqn
\ScalM(z+\varepsilon z;x+\varepsilon\xi,y+\varepsilon\xi)&=&\ScalM(z;x,y)+\left[\ScalM,\varepsilon z\Conn_z+\varepsilon\xi^{\nu}\Conn_{\nu}\right]+\varepsilon\mathrm{Tr}\;\gamma(x,y;u,v)\cdot\ScalM(v,u)\nonumber\\
&+&\varepsilon\mathrm{Tr}\;\gamma_{\mu}(x,y;u,v)\cdot\ConnM^{\mu}(v,u)+O(\varepsilon^2) \label{CS1}
\eeqn
\beqn
\ConnM^{\mu}(z+\varepsilon z;x+\varepsilon\xi,y+\varepsilon\xi) &=& \ConnM^{\mu}(z;x,y)-\left[\varepsilon zW_z+\varepsilon\xi^{\nu}\Conn_{\nu},\ConnM^{\mu}\right]-\varepsilon\frac{N}{2}\alp^{\mu}+\varepsilon\mathrm{Tr}\;\gamma^{\mu}(x,y;u,v)\cdot\ScalM(v,u)\nonumber\\
&+&\varepsilon\mathrm{Tr}\;{{\gamma}^{\mu}}_{\nu}(x,y;u,v)\cdot\ConnM^{\nu}(v,u)+O(\varepsilon^2) \label{CS2}
\eeqn
where we have defined the gamma functions, whose explicit expressions can be found in Appendix \ref{RGappendix}. We note that they have the properties
\beq
\gamma(x,y;u,v) = \frac{\delta \beta^{(\Scal)}(u,v)}{\delta \Scal(x,y)},\quad\quad {{\gamma}^{\mu}}_{\nu}(x,y;u,v) = \frac{\delta \beta^{(\Conn)}_{\nu}(u,v)}{\delta \Cont_{\mu}(x,y)} \label{gammafunctions1}
\eeq
\beq
\gamma_{\mu}(x,y;u,v) = \frac{\delta \beta^{(\Conn)}_{\mu}(u,v)}{\delta \Scal(x,y)}=\frac{\delta \beta^{(\Scal)}(u,v)}{\delta \Cont^{\mu}(x,y)}
\label{gammafunctions2}
\eeq
which will play an important role in the next section.
Note that $\ScalM$ and $\ConnM^{\mu}$ transform tensorially under $O(L_2)$. We denote the bulk extensions of the momenta $\ScalM$ and $\ConnM^{\mu}$ as $\cScalM$ and $\cConnM^{\mu}$ respectively. Comparing the terms proportional to $\varepsilon$ on both sides of equations \eqref{CS1} and \eqref{CS2}, we obtain in the limit $\varepsilon \mapsto 0$
\beq
[\cD_{\ue^{(0)}_z},\cScalM](x,y)=\left\{\mathrm{Tr}\;\gamma(x,y;u,v)\cdot\cScalM(v,u)+\mathrm{Tr}\;\gamma_{\mu}(x,y;u,v)\cdot\cConnM^{\mu}(v,u)\right\}\label{MajCS1}
\eeq
\beq
\left[\cD_{\ue^{(0)}_z},\cConnM^{\mu}\right](x,y) = \left\{-\frac{N}{2}\alp^{\mu}(x,y)+\mathrm{Tr}\;\gamma^{\mu}(x,y;u,v)\cdot\cScalM(v,u)+\mathrm{Tr}\;{{\gamma}^{\mu}}_{\nu}(x,y;u,v)\cdot\cConnM^{\nu}(v,u)\right\} \label{MajCS2}
\eeq
where as before $\cD = \bd+\cConn$. 

Finally, in preparation for forthcoming discussions, we also write down the Ward identity for RG transformations:
\begin{align}
	\frac{\partial}{\partial z}Z
	&= - \mathrm{Tr}\left\{ \left( \left[ \Scal , \Conn_{\ue^{(0)}_z} \right]_\bl + \beta^{(\Scal)} \right) \bl \frac{\delta}{\delta \Scal}
	     + \left( \left[ P_{F;\mu} + \Conn_\mu , \Conn_{\ue^{(0)}_z}\right]_\bl + \beta^{(\Conn)}_{\mu} \right) \bl \frac{\delta}{\delta \Conn_\mu} \right\} Z \nonumber\\
	     \label{RGWard2} &\quad
	     + \frac{N}{2}\mathrm{Tr}\left\{\alp^\mu \bl \Cont_\mu + \alp^z \bl \Cont_{\ue^{(0)}_z}\right\}Z
\end{align}
This is just an infinitesimal version of equation \eqref{step2pt2}. Note that by $\frac{\partial}{\partial z}Z$ we mean the partial derivative with respect to $z$, keeping the sources fixed.  In the last line of \eqref{RGWard2}, we have taken into account that the potential $\pot$ is also modified as we move into the bulk.  The $\alp_z$ appearing in the final term denotes a possible $CO(L_2)$ Weyl anomaly, with the notation chosen suggestively (the $\alp_\mu \Cont^\mu$ comes directly from the transformation of the determinant).  This notation is discussed further in Appendix \ref{RGappendix}.

We will see next that, from a holographic point of view, equation \eqref{RGWard2} can be interpreted as the \emph{Hamilton-Jacobi} equation \cite{deBoer:1999xf}. With this interpretation, the RG equations and the Callan-Symanzik equations then turn out to be Hamilton equations of motion.

\subsection{Holography as Hamilton-Jacobi} \label{sec:hamiltonjacobi}

Let us now switch to a holographic perspective and consider the free-field Majorana fermions as living on the conformal boundary of a $d+1$-dimensional, asymptotically $AdS$  spacetime $(M_{d+1},G)$. 
Corresponding to the operators (whose vevs are) $\ScalM$ and $\ConnM^{\mu}$ sourced by $\Scal$ and $\Conn_{\mu}$ in the boundary field theory, we usually think in terms of bulk fields whose dynamics relate them. In Ref. \cite{deBoer:1999xf}, it was proposed that the nature of holographic renormalization is encapsulated in the relationship 
\beq
Z[M, z, \Scal(z),\Conn_{\mu}(z)] = e^{iS_{HJ}[z,\Scal(z),\Conn_{\mu}(z)]}
\eeq 
where $S_{HJ}$ is interpreted as the Hamilton-Jacobi functional associated with given radial dynamics. Given such bulk dynamics, this coincides with the on-shell action written as a functional of boundary values of the fields.

Thus at the heart of the holographic principle is the \emph{Lifshitz}\footnote{Here we are using language analogous to Lifshitz field theories, whose vacuum wave functional is given by the exponential of a spatial CFT action.} property - the generating functional in the boundary is a wavefunctional in the bulk, from the point of view of radial quantization in which the `time parameter' is the radial coordinate $z$ \cite{deBoer:1999xf}. From this point of view, the connected vacuum expectation values $\ScalM$ and $\ConnM^{\mu}$ given by
\beq
\ScalM = \frac{\delta S_{HJ}}{\delta \Scal},\;\;\;\ConnM^{\mu}=\frac{\delta S_{HJ}}{\delta\Conn_{\mu}}
\eeq
can then be thought of as the boundary values of momenta $\cScalM$ and $\cConnM^{\mu}$ conjugate to bulk fields $\cScal$ and $\cConn_{\mu}$ respectively. 

Now, from the point of view of the boundary field theory, we are not given directly bulk dynamics, but we are given $S_{HJ}$, and one could attempt to reconstruct a choice of bulk dynamics that reproduces it. We wish to identify the bulk theory with a (classical) higher spin theory, but it is not clear if a local action exists. 

We can however proceed further. We observe that 
the RG Ward identity \eqref{RGWard2} takes the form of the Hamilton-Jacobi equation
\beq\label{HamiltonJacobi}
\frac{\partial}{\partial z}S_{HJ} = -\mathcal{H}
\eeq
with the Hamiltonian\footnote{This should be distinguished from the Hamiltonian constraint of gravitational theories.}
\beqn\label{RGHamiltonian}
\mathcal{H}&=& -\mathrm{Tr} \left\{\left(\left[\cScal,\cConn_{\ue^{(0)}_z}\right]_\bl +\beta^{(\cScal)}\right)\bl\cScalM+\left(\left[P_{F;\mu}+\cConn_{\mu},\cConn_{\ue^{(0)}_z}\right]_\bl+{\beta}^{(\cConn)}_{\mu}\right)\bl\cConnM^{\mu} \right\}\nonumber\\
&-&\frac{N}{2}\mathrm{Tr}\left\{ \left( \alp^{\mu}\bl\cCont_{\mu}+ \alp^z \bl \cCont_{\ue^{(0)}_z} \right) \right\}
\eeqn
It is straightforward to check that the $dz$ components of the RG equations \eqref{MajRG1}, \eqref{MajRG2}, and the Callan-Symanzik equations  \eqref{MajCS1}, \eqref{MajCS2} derived in the previous section are precisely the Hamilton equations of motion 
\beq
d_z\cScal = \frac{\delta\mathcal{H}}{\delta \cScalM},\quad 
d_z\cConn_{\mu} = \frac{\delta\mathcal{H}}{\delta \cConnM^{\mu}},\quad 
d_z\cScalM = -\frac{\delta\mathcal{H}}{\delta \cScal},\quad 
d_z\cConnM^{\mu} = -\frac{\delta\mathcal{H}}{\delta \cConn_{\mu}},
\eeq 
Note that equations \eqref{gammafunctions1} and \eqref{gammafunctions2} are sort of integrability conditions in making this Hamiltonian formalism work. Note also, that $\cConn_z$ has no dynamics of its own; the partition function does not depend on it, and thus its conjugate momentum is zero. Thus $\cConn_z$ is a Lagrange multiplier, which enforces the Ward identity associated with $O(L_2)$. 

Additionally of course, we have the transverse equations of motion, i.e. the $dx^{\mu}$ components of equations \eqref{MajRG1}, \eqref{MajRG2}. If we implement these constraints by introducing additional non-dynamical Lagrange multipliers ${\cal Q}^\mu$, $\cal Q^{\mu\nu}$ 
\beq
{\cal H}_{constraint}=
-\mathrm{Tr} \Big\{
\left(\left[\cD_{\mu},\cScal\right]-\beta^{(\cScal)}_{\mu}\right)
\bl {\cal Q}^{\mu}
+\left(\cF_{\mu\nu}-\beta^{(\cConn)}_{\mu\nu}\right)
\bl{\cal Q}^{\mu\nu}
\Big\}
\eeq
then the full Hamiltonian $(\mathcal{H} + \mathcal{H}_{constraint})$ might be taken to give rise to an `action' (written in terms of phase space variables)
\beq
I=\int dz\;\mathrm{Tr}\left\{ \cScalM^I\cdot \left(\left[ \cD_I,\cScal\right]-\Beta^{(\cScal)}_{I}\right)+\cScalM^{IJ}\cdot\left(\cF_{IJ}-\Beta^{(\cConn)}_{IJ}\right)-\frac{N}{2}\alp_I\cdot \widehat{\cConn}^I\right\}
\eeq
where we have collected $\cScalM$ and ${\cal Q}^\mu$ into the components of a 1-form $\cScalM^I$ and $\cScalM^\mu$ and $\cal Q^{\mu\nu}$ into a 2-form $\cScalM^{IJ}$. The equations of motion derived from this action are equivalent to our RG and Callan-Symanzik equations provided we gauge-fix all the Lagrange multipliers to zero. This sort of action has been proposed before in several contexts \cite{Dolan:1994eg,Doroud:2011xs,Boulanger:2011dd,Boulanger:2012bj}. Since the Hamiltonian is linear in momenta, we are not free to pass back and forth between Hamiltonian and Lagrangian formulations in the usual way. 

Two important comments are in order here:

(i) \textbf{First order v/s Second order}: It is clearly very important here that the Hamiltonian $\mathcal{H}$ \eqref{RGHamiltonian} is linear in momenta - this means that the RG equations do not involve momenta, and can be solved (in principle) on their own, without reference to the conjugate momenta. Subsequently, we may solve the Callan-Symanzik equations to obtain the radial evolution of momenta. 

Thus in this sense, the RG equations are intrinsically \emph{first-order} in nature. As was mentioned before, this is a special property of vector models, namely that it is possible to truncate the RG flow out of the free fixed point to single-trace operators. It is the fact that this system is closed (other operators are not sourced by the flow) that corresponds to the Hamiltonian being linear in momenta. Of course, in most other field theories with interactions such a truncation is not possible. For instance in matrix models, the generation of multi-trace operators makes the Hamiltonian for radial evolution (RG flow) quadratic in momenta, thus intertwining the RG equations with Callan-Symanzik equations. It is useful to compare these results with those of Refs. \cite{Lee:2009ij},\cite{Lee:2010ub},\cite{Lee:2013dln}. Indeed, from that point of view, the single trace vector models are a special case where because of the absence of interactions, no bulk dynamics is generated (because multi-trace operators are not generated and thus do not need to be disentangled at each scale by introducing new degrees of freedom). Further discussion of interacting theories can be found in Section \ref{Sec:interactions}.

(ii) \textbf{The two point function}: As a check of the consistency of our results, we compute the 2-point function of the elementary field. Given the formalism, we can't do this completely, but we can extract its $O(N)$ trace
\beq
S_\alpha{}^\beta(y,x)\equiv\langle \psi^m_\alpha(y)\tpsi^{m,\beta}(x)\rangle= -\ScalM(x,y)\delta_\alpha^\beta-\ConnM_\mu(x,y)(\gamma^\mu)_\alpha{}^\beta.
\eeq
At the free fixed point $A=0=\widehat{W}$, the Callan-Symanzik equations simplify to
 \beqn
\left[ {\cal D}^{(0)},\Pi_A\right](x,y)&=&0\\
\left[ {\cal D}^{(0)},\Pi^\mu\right](x,y)&=& -\frac{N}{2}\Delta^\mu(x,y)dz.\label{freeRG}
\eeqn
Given the definition of $\Delta_\mu$ in eq. (\ref {defDelta}), we then find
\beq
\gamma_\mu\Delta^\mu(x,y)=-z\pa_z (i\slashed{D}_F^{(0)})^{-1}
\eeq
and thus eq. (\ref{freeRG}) integrates to
\beq
S=iN(\slashed{D}_F^{(0)})^{-1}.
\eeq
We obtain the expected inverse Dirac operator, and the factor of $N$ is expected since we computed the $O(N)$-trace. Thus we see the significance of the $\Delta_\mu$ term in the $\beta$-functions. 

(iii) \textbf{Structure of the beta functions}: What role do the $\beta$-function terms which appear in the Hamiltonian, and RG equations play? To address this, it is useful to make contact with conventional understanding of RG within the context of conformal perturbation theory. Let $\mathcal{O}_i$ be a complete set of operators at a given fixed point, and we label by $\lambda^{i}$, the corresponding coordinates on the coupling-constant space of deformations away from the fixed point
\beq
S_{perturb.} = \sum_{i} \int d^dx \lambda^{i}(x)\mathcal{O}_{i}(x)
\eeq
In our case of course, all the $\lambda_i$ s are contained in the tensorial bilocal sources $\Scal$ and $\Cont_{\mu}$. From conformal perturbation theory, the beta functions for renormalization group flow take the form
\beq
d_z\lambda^i \equiv \beta_0^i = {\Gamma^i}_j\lambda^j + {\mathcal{C}^i}_{jk}\lambda^j\lambda^k+\cdots
\eeq 
where ${\Gamma^i}_j$ and ${\mathcal{C}^i}_{jk}$ are constants associated with the fixed point. One might in certain situations, find it natural to combine the ${\Gamma^i}_j\lambda^j$ term with $d_z\lambda^i$ to define a ``covariant derivative'' $D_z\lambda^i=(d_z\lambda^i-{\Gamma^i}_j\lambda^j)$. Indeed, this is precisely the case in our RG equations \eqref{MajRG1}, \eqref{MajRG2}, and consequently, our tensorial beta functions \eqref{beta1}, \eqref{beta2} must schematically be compared to 
\beq
D_z\lambda^i\equiv \beta^i = {\mathcal{C}^i}_{jk}\lambda^j\lambda^k+\cdots
\eeq 
Now at the free fixed point, the constants ${\mathcal{C}^i}_{jk}$ are closely related to the OPE coefficients
\beq
\mathcal{O}_i\mathcal{O}_j \sim \sum_k{c^k}_{ij}\mathcal{O}_k 
\eeq
Indeed, given that the free-field OPE essentially involves contracting elementary fields between the two operators, a closer look at the Polchinski ERG formalism reveals
\beq
{\mathcal{C}^i}_{jk} = Md_M {c^i}_{jk}
\eeq
Thus, by extracting the coefficients ${\mathcal{C}^i}_{jk}$ from the beta functions, it is straightforward to read off the OPE coefficients ${c^i}_{jk}$.\footnote{For instance, by looking at the $\Scal\bl\alp_{\mu}\bl\Scal$ term in $\beta^{(\Conn)}_{\mu}$, we obtain
\beq
2{c^{\Conn_{\mu}(x,y)}}_{\Scal(u,v)\;\Scal(w,z)}= \delta(x-z)G_{\mu}(w,u)\delta(v-y)+\delta(x-u)G_{\mu}(v,z)\delta(w-y)\nonumber
\eeq  
where $\gamma^{\mu}G_{\mu}(x,y)$ is the free-fermion Green function, with $\alp_{\mu}=Md_MG_{\mu}$. The other coefficients may be computed similarly.}  

This discussion sheds new light on the structure of our beta functions -- from the field theory point of view, the beta functions encode information about the OPE coefficients, and hence the 3-point functions of the free-fermion CFT. On the other hand, from the holographic point of view, 3-point functions of the CFT are dual to 3-point tree level scattering amplitudes in the bulk. \emph{We thus conclude that the beta function terms in our RG equations encapsulate cubic interactions in the bulk}. Detailed computations of tree level 3-point scattering amplitudes in Vasiliev higher spin theory in $AdS_4$ have been carried out in \cite{Giombi:2009wh,Giombi:2012ms}, and were found to be in agreement with the 3-point functions of the CFT, in the case of the bosonic $O(N)$ vector model. 

\section{The Infinite Jet Bundle}\label{sec: jetbundles}

Although we have talked about general bilocal symmetries thus far, in order to make more direct contact with higher spin theory, it is convenient to introduce a \emph{quasi-local} expansion for the sources
\beq
\cScal(z;x,y) \simeq \sum_{s=0}^{\infty}\cScal^{a_1\cdots a_s}(z,x)\;\pa^{(x)}_{a_1}\cdots \pa^{(x)}_{a_s}\delta^d(x-y) \label{QLE1}
\eeq
\beq
\cConn_{I}(z;x,y) \simeq \sum_{s=1}^{\infty}{\cConn_{I}}^{a_1\cdots a_{s-1}}(z,x)\;\pa^{(x)}_{a_1}\cdots \pa^{(x)}_{a_{s-1}}\delta^d(x-y). \label{QLE2}
\eeq
From the field theory point of view, this means we source all \emph{local} single trace operators with no prejudice towards the number of derivatives they contain. In this section, we will try to clarify the meaning of the above quasi-local expansion. More importantly, we wish to make mathematically precise the sense in which the $CO(L_2)$ symmetry discussed previously is a gauge symmetry and $\cConn$ is a connection. Naively, such a gauge-theoretic interpretation of our bilocal symmetries would require a vector bundle over spacetime, with the fiber being the space of all $L_2$ functions over spacetime. As we will see shortly, this leads us naturally to the idea of jet bundles. 

Before we get into the details, we outline the basic intuition behind the following construction. In physics, a connection is usually thought of as a Lie-algebra valued 1-form $W = W_{\mu}^aT^a dx^{\mu}$, which gives us a covariant derivative while acting on fields charged under the corresponding gauge symmetry. In order to truly interpret our 1-form $\cConn$ as a $CO(L_2)$ connection, we need to cast it in this language. Indeed, equation \eqref{QLE2} can roughly be thought of as 
\beq
\cConn(z;x,y) = \sum_{s=1}^{\infty}\cConn^{a_1\cdots a_{s-1}}(z,x)\mathbb{T}_{a_1\cdots a_{s-1}}(x,y),\;\;\;\; \mathbb{T}_{a_1\cdots a_{s-1}}(x,y)\simeq \partial^{(x)}_{a_1}\cdots \partial^{(x)}_{a_{s-1}}\delta^d(x-y)
\eeq
In order to interpret the $\mathbb{T}_{a_1\cdots a_{s-1}}$ as a matrix (not in the functional sense) acting on the elementary fields, it is useful to think of the field  $\psi^a$ and all its derivatives at a point, as forming a vector
\beq
\left(\psi^m(x),\frac{\partial\psi^m}{\partial x^{\mu}}(x),\frac{\partial^2\psi^m}{\partial x^{\mu}\partial x^{\nu}}(x),\cdots\right). \label{pro}
\eeq
$\mathbb{T}_{a_1\cdots a_{s-1}}$ can then be thought of as a matrix, acting linearly on this vector. The corresponding gauge symmetry then locally (i.e., in a spacetime-dependent way) mixes the various derivatives of $\psi^m$ pointwise and linearly. In mathematics, this simple idea fits in with the notion of a \emph{jet bundle}. The vector \eqref{pro} is called a {\it jet} corresponding to $\psi^m$, and $\cConn$ is naturally interpreted as a connection on the jet bundle. The following section introduces mathematical details of this construction. Less mathematically-minded readers may skip forward to section \ref{sec:COL2rev} keeping in mind that the jet bundle construction allows us to think about sets of derivatives of fields in `vector bundle' terms.

\subsection{Mathematical Preliminaries} \label{mathprelims}

For any given value of $z$, the elementary fields $\psi^m(x)$ in the field theory are sections of the Majorana bundle\footnote{More precisely, they are sections of the product of the Majorana bundle associated with the spin bundle with a trivial $\re^N$ bundle, where $N$ is the number of flavors of fermions. The spin and $O(N)$ indices are largely spectators in the geometric construction that we are describing here, and thus the construction applies equally well to any sort of field.} $E$ over $M_d=\re^{d}$, which we denote by $\pi:E\mapsto M_d$. We will label by $\Gamma(E)$ the space of all $C^{\infty}$ sections of $E$. Corresponding to $E$, there exists the \emph{infinite jet bundle} over $M_d$
\beq
\pi_{\infty}: J^{\infty}(E)\mapsto M_d
\eeq
which is defined as follow: two sections $\psi^m(x)$ and $\chi^m(x)$ of $E$ are said to have the same $r$th jet at a point $x \in M_{d}$ if 
\beq
\left.\frac{\pa^k}{\pa x^{a_1}\cdots \pa x^{a_k}}\psi^m\right|_x = \left.\frac{\pa^k}{\pa x^{a_1}\cdots \pa x^{a_k}}\chi^m\right|_x,\;\;\; 0\leq k \leq r
\eeq
For any given section $\psi^m(x)$ of $E$, the $r$th jet of $\psi^m$ at $x$, denoted by $j^r_x\psi$, is the equivalence class of all sections which have the same $r$th jet at $x$ as $\psi^m$.  The $r$th \emph{jet bundle} $\pi_r: J^r(E) \mapsto M_d$ of $E$ over $M_d$ is then defined by
\beq
J^r(E) = \left\{j^r_x\psi\;:\; \forall x \in M_d, \psi \in \Gamma(E)\right\}
\eeq
with the natural projection $\pi_r: j^r_x \psi \mapsto x$. The \emph{infinite jet bundle} $J^{\infty}(E)$ of $E$ is defined as above, with $r \to\infty$. Given a section $\psi^m(x)$ of $E$, we can naturally construct a section $j^{\infty}\psi^m(x)$ of $J^{\infty}(E)$ by taking its infinite jet at every point $x$. This is called the \emph{prolongation} map
\beq
j^{\infty}: \Gamma(E) \mapsto \Gamma(J^{\infty}(E))
\eeq
In simple terms, the prolongation map sends
\beq
\Gamma(E) \ni \psi^m(x) \mapsto \left(\psi^m(x), \frac{\pa \psi^m}{\partial x^{a_1}}(x), \frac{\partial^2 \psi^m}{\pa x^{a_1}\pa x^{a_2}}(x),\cdots \right) \in \Gamma(J^{\infty}(E)).
\eeq
The important point is that a differential operator can be thought of as a section of the \emph{Endormorphism bundle} $\mathrm{End}(J^{\infty}(E))$ of $J^{\infty}(E)$, i.e. it is simply a local linear transformation when thought of as acting on sections of the jet bundle. For instance, the derivative operator $\frac{\pa}{\pa x^\mu}$ can loosely be thought of as the matrix (in terms of a local trivialization)
\beq
\fr{P}_{\mu} = \left(\begin{matrix} \mathbf{0} &  &\mathbf{1} & &\mathbf{0}& &\mathbf{0} & &\cdots\\ & & & & & & & & \\\mathbf{0} & & \mathbf{0} & & \mathbf{1} & & \mathbf{0} & & \cdots \\ & & & & & & & & \\ \mathbf{0} & & \mathbf{0} & & \mathbf{0} & & \mathbf{1} & & \cdots\\ \vdots  & &\vdots  & & \vdots & &   \end{matrix}\right)\label{P}
\eeq
with each entry corresponding to a map between tensors of different ranks. In more precise notation, $\fr{P}_{\mu}$ is a section of $\mathrm{End}(J^{\infty}(E))$. Acting on a vector $j^{\infty}_x\psi^m$ at $x$, it may be defined as the push-forward of the derivative operator:
\beq
\left(\fr{P}_{\mu}\cdot j_x^{\infty}\psi^m\right)(x) = j^{\infty}_x\left(\partial_{\mu}\psi^m\right)(x)
\eeq
or in terms of a commuting diagram
\begin{displaymath}
\xymatrix{\Gamma(E) \ar[r]^{\pa_{\mu}} \ar[d]_{j_x^{\infty}} & \Gamma(E) \ar[d]^{j_x^{\infty}} \\ J_x^{\infty}(E) \ar[r]_{\fr{P}_{\mu}} & J_x^{\infty}(E) }
\end{displaymath}
where, by $J_x^{\infty}(E)$ we mean the fiber of the infinite jet bundle over $x$. Similarly, we may also construct the operator $\fr{X}^{\mu}$ (again as a section of $\mathrm{End}(J^{\infty}(E))$), acting on the vector $j^{\infty}_x\psi^m$ at $x$ as
\beq
\left(\fr{X}^{\mu}\cdot j_x^{\infty}\psi^m\right)(x) = j_x^{\infty}\left(x^{\mu}\psi^m\right)(x)
\eeq
In other words, $\fr{X}^{\mu}$ is the push-forward of multiplication by $x^{\mu}$
\begin{displaymath}
\xymatrix{\Gamma(E) \ar[r]^{x_{\mu}} \ar[d]_{j_x^{\infty}} & \Gamma(E) \ar[d]^{j_x^{\infty}} \\ J_x^{\infty}(E) \ar[r]_{\fr{X}_{\mu}} & J_x^{\infty}(E)} 
\end{displaymath}
Going further, we can use $\fr{X}$ and $\fr{P}$ to construct more complicated matrices, such as generators of $\mathfrak{so}(2,d)$
\begin{align}\label{so(2,d)gen}
\fr{M}_{ab}&=\fr{X}_a\fr{P}_b-\fr{P}_a\fr{X}_b\nonumber\\
\fr{D} &= \fr{X}^a\fr{P}_a\\
\fr{K}_{a} &= \fr{X}^2\fr{P}_a-2\fr{X}_a\fr{X}_b\fr{P}^b\nonumber
\end{align}
which may easily be shown to satisfy the appropriate commutation relations. It is also convenient to introduce a bilinear form on the fibres of $J^{\infty}(E)$ which, intuitively speaking, we want to look like
\beq
\langle\cdot, \cdot \rangle  = \left(\begin{matrix} \mathbf{1} &  &\mathbf{0} & &\mathbf{0} & &\mathbf{0} & &\cdots\\ & & & & & & & & \\\mathbf{0} & & \mathbf{0} & & \mathbf{0} & & \mathbf{0} & & \cdots \\ & & & & & & & & \\ \mathbf{0} & & \mathbf{0} & & \mathbf{0} & & \mathbf{0} & & \cdots\\ \vdots  & &\vdots  & & \vdots & &   \end{matrix}\right)\otimes \epsilon_{\alpha\beta}\otimes \delta_{mn}
\eeq
where $\epsilon_{\alpha\beta}$ and $\delta_{mn}$ are the metrics for spinor and $O(N)$ indices respectively. More precisely then, we define $\langle \cdot, \cdot \rangle$ as
\beq
\langle j^{\infty}_x\psi^m, j^{\infty}_x\chi^n\rangle(x) = \delta_{mn}\widetilde{\psi}^m(x)\chi^n(x)
\eeq
With this, we naturally get an \emph{inner product} $\langle\cdot,\cdot \rangle_{\Gamma(J^{\infty}(E))}$ on sections of $J^{\infty}(E)$
\beq\label{metric}
\langle \Phi^m, \Psi^n\rangle_{\Gamma(J^{\infty}(E))} = \int_{M_d}d^dx\sqrt{g^{(0)}(x)}\;\langle \Phi^m(x),\Psi^n(x)\rangle
\eeq
where $\Phi,\Psi \in \Gamma(J^{\infty}(E))$, and we have made the (metric) measure on spacetime explicit. 
The point of choosing this inner product of course, is that \emph{on prolongations}, it agrees with the standard inner product on $\Gamma(E)$, namely 
\beq\label{metric}
\langle j^{\infty}\psi^m, j^{\infty}\chi^n\rangle_{\Gamma(J^{\infty}(E))}=\langle \psi^m,\chi^n\rangle_{\Gamma(E)}=\int_{M_d} d^dx\;\sqrt{g^{(0)}(x)}\;\delta_{mn}\widetilde{\psi}^m(x)\chi^n(x)
\eeq 
We can express this succinctly in terms of a commutative diagram as follows:
\begin{displaymath}
\xymatrix{\Gamma(E)\times \Gamma(E) \ar[r]^{\;\;\;\;\;\langle \cdot,\cdot \rangle_{\Gamma(E)}} \ar[d]_{j^{\infty}} & \mathbb{R}  \\ \Gamma(J^{\infty}(E))\times \Gamma(J^{\infty}(E)) \ar[ur]_{\;\;\;\;\;\langle \cdot,\cdot \rangle_{\Gamma(J^{\infty}(E))}} &}
\end{displaymath}

\subsection{$CO(L_2)$ in the Jet Bundle Language}  \label{sec:COL2rev}
Now we come to the crucial point of all this discussion: \emph{We interpret the (quasi-local) 1-form source $W(z)=\Conn_{\mu}dx^{\mu}$ (on a given $z$-slice) as a connection\footnote{Recall that a connection on a vector bundle $\pi:V\mapsto M$ over $M$ is a section of $T^{*}M\otimes \mathrm{End}(V)$, i.e. a 1-form on $M$ taking values in the endomorphisms of fibers.} on the infinite jet bundle $J^{\infty}(E)$ at $z$, and $\Scal(z)$ as a section of $\mathrm{End}\;(J^{\infty}(E))$. Further, by extending $J^{\infty}(E)$ trivially in the $z$ direction\footnote{More precisely, we mean that we take the jet bundle of $E$ and extend that to the bulk. This would not be the same as extending $E$ to the bulk and taking its jet bundle. In other words, the Taylor expansions are in the transverse space only. Also, since we are always in a local patch (namely the Poincar\'e patch), there are no topological obstructions to extending the jet bundle into the bulk.} to a bundle $J_{bulk}^{\infty}(E)$ over the bulk $M_{d+1}$, the bulk 1-form $\cConn=\cConn_Idx^I$ becomes a connection over $J_{bulk}^{\infty}(E)$, while the 0-form $\cScal$ is a section of $\mathrm{End}\;(J^{\infty}_{bulk}(E))$.} 

Let us label a basis of sections (or a local trivialization) of $\mathrm{End}(J^{\infty}(E))$ by $\{\fr{T}_{\uM}\}$. We thus interprete  and generalize the quasi-local expansions \eqref{QLE1}, \eqref{QLE2} in concrete terms as 
\beq
\cScal(z;x,y)\mapsto \sum_{\uM}\cScal^{\uM}(z,x)\;\fr{T}_{\uM},\;\;\;\cConn_{I}(z;x,y) \mapsto \sum_{\uM}\cConn_{I}^{\uM}(z,
x)\;\fr{T}_{\uM}.
\eeq
Note the significance of this reinterpretation - we have translated bilocal kernels mapping functions (on spacetime) to functions, into local operators mapping sections (of a vector bundle) to sections. While this might seem like a technical point, it profoundly facilitates the identification of gauge theory structure in the $O(L_2)$ and $CO(L_2)$ symmetries. 

Indeed, by requiring that the inner product on $\Gamma(J^{\infty}(E))$ (see equation \eqref{metric}) be preserved, we may reduce the structure group down to generators $\epsilon=\epsilon^{\uM}\fr{T}_{\uM}$ which satisfy the condition 
\beq\label{jetol2}
\left\langle \Phi^a,\epsilon\; \Psi^b\right\rangle_{\Gamma(J^{\infty}(E))}+\left\langle\epsilon\; \Phi^a,\Psi^b\right\rangle_{\Gamma(J^{\infty}(E))} = 0.
\eeq
for all $\Psi,\;\Phi\in \Gamma(J^{\infty}(E)$. This is analogous to the familiar idea of reducing the structure group of (for instance) the tangent bundle of a manifold from $GL(n)$ to $O(n)$ by picking a metric on it. Equation \eqref{jetol2} is what we called the $O(L_2)$ condition previously (see equation \eqref{OL2condition}), and may be suggestively written as
\beq
``\;\;\;\epsilon+\epsilon^T = 0\;\;\;"
\eeq
The space of all such generators forms a Lie-algebra (with the bracket being the commutator), which we may refer to as $\mathfrak{o}(L_2)$. Enlarging to $CO(L_2)$ amounts to preserving the inner product up to a local scale transformation $g^{(0)}\mapsto \lambda^2g^{(0)}$. The corresponding Lie-algebra may be referred to as $\mathfrak{co}(L_2)$. It is an easy exercise to check that the $\mathfrak{so}(2,d)$ generators in equation \eqref{so(2,d)gen} all belong to $\mathfrak{co}(L_2)$.

As was pointed out above, the main utility of the jet-bundle formalism is that it allows us to cast our previous discussion of symmetries in the language of vector bundles. For example, an infinitesimal $O(L_2)$ or $CO(L_2)$ gauge parameter $\epsilon(z;x,y)$ is now to be replaced by $\epsilon = \epsilon^{\uM}(z,x)\fr{T}_{\uM}$ in $\mathfrak{o}(L_2)$ or $\mathfrak{co}(L_2)$ respectively. 
The action of the gauge symmetries on $\cScal$ and $\cConn$ 
\beq
\delta\cScal = \left[\epsilon, \cScal\right],\;\; \delta\cConn = \bd\epsilon+\left[\cConn,\epsilon\right]
\eeq
remains the same, with the commutators appropriately reinterpreted as matrix commutators of the $\fr{T}_{\alpha}$'s. 

As before, a natural choice for the flat background connection $\cConn^{(0)}$ is the $AdS_{d+1}$ connection (equation \eqref{AdSconnection}), which in the present language takes the form
\beq
\cConn^{(0)}(z,x) = -\frac{dz}{z}\fr{D}+\frac{dx^{a}}{z}\fr{P}_a
\eeq
From the field theory point of view, $\cConn^{(0)}$ encodes the pure-gauge renormalization group flow of the free-fixed point. The  ``global symmetries'' of the free-fixed point (i.e. symmetry transformations which leave the unperturbed free CFT invariant) are therefore naturally identified with the maximal Lie-subalgebra within $\mathfrak{co}(L_2)$ comprising of elements $\epsilon^{(0)}$ which preserve $\cConn^{(0)}$, namely
\beq
\delta \cConn^{(0)} = \bd\epsilon^{(0)}+\left[\cConn^{(0)}, \epsilon^{(0)}\right]=0
\eeq
It is not hard to obtain a basis for this subalgebra:
\beqn\label{subalgebra}
\fr{T}_d&=& \fr{D}\nonumber\\
\fr{T}_a&=& \fr{P}_a,\nonumber\\
\fr{T}_{a,b} &=& \fr{M}_{a,b},\nonumber\\
\fr{T}_{a_1\cdots a_{s-1};b_1\cdots b_t} &=& \left(\fr{M}_{a_1,b_1}\cdots \fr{M}_{a_t,b_t}\fr{P}_{a_{t+1}}\cdots\fr{P}_{a_{s-1}}\right)_{W} 
\eeqn
with $s=2,4,6\cdots$. (Also, the boundary indices $a,b$ run over 0 to $d-1$, and the radial direction is labelled as $d$.) The subscript $W$ indicates that the product is Weyl-ordered, which is essential in order for the element to lie within $\mathfrak{co}(L_2)$. For instance, the first few linear combinations which leave the background connection invariant are
\beq
\frac{1}{z}\fr{T}_a, \;\;\;\;\left(\fr{T}_{a,b}-\frac{1}{z}x_{[a}\fr{T}_{b]}\right),\;\;\;\;\left(\fr{T}_d-\frac{1}{z}x^a\fr{T}_a\right), \cdots
\eeq
From a physics point of view, it is natural to project down to this subalgebra by only considering gauge transformations of the form
\beq
\epsilon(z,x) = \sum_{s,t} \epsilon^{a_1\cdots a_{s-1};b_1\cdots b_t}(z,x)\fr{T}_{a_1\cdots a_{s-1};b_1\cdots b_t}
\eeq
On the field theory side, this corresponds to ``gauging'' the global symmetries of the free-fixed point. The connection may be then taken to be
\beq
\cConn(z,x) = \frac{dz}{z}\fr{T}_d+\sum_{s,t}\left(dz\cConn_{z}^{a_1\cdots a_{s-1};b_1\cdots b_t}(z,x)\fr{T}_{a_1\cdots a_{s-1};b_1\cdots b_t}+dx^{\mu}\cConn_{\mu}^{a_1\cdots a_{s-1};b_1\cdots b_t}(z,x)\fr{T}_{a_1\cdots a_{s-1};b_1\cdots b_t}\right)
\eeq
Note that the fields $\cConn^{a_1\cdots a_{s-1};b_1\cdots b_t}$ are in 
\beq
\begin{Young}
&&&&&\cr
&&&\cr
\end{Young}\nonumber
\eeq
representations of the boundary Lorentz group $O(1,d-1)$, and can be thought of as sourcing higher-spin currents in the boundary theory. Similarly, the 0-form $\cScal$ takes the form
\beq
\cScal(z,x) = \sum_{s,t}\cScal^{a_1\cdots a_s;b_1\cdots b_t}(z,x)\fr{T}_{a_1\cdots a_{s};b_1\cdots b_t}
\eeq
where $s$ may be taken to be even because the Majorana theory is only sensitive to the symmetric part of $\cScal$. 

In fact, let us introduce a specific representation of the symmetry generators, that will allow us to make direct contact with the formalism of Vasiliev (some background discussion of the basic structure of Vasiliev's higher spin theory can be found in Appendix B, which we have included for completeness.)
Indeed, a representation for the generators in equation \eqref{subalgebra} is obtained by introducing the variables $Y^A_i$ (where $A=-1,0,\cdots, d$ are $SO(2,d)$ vector indices while $i=1,2$ are $sp(2)$ indices), endowed with the \emph{star-product}
\beq
Y^A_i\star Y^B_j = Y^A_iY^B_j+\frac{1}{2}\eta^{AB}\epsilon_{ij}
\eeq
(Alternatively, and perhaps more appropriate to the case at hand, one can introduce a representation in terms of twistor variables. Here, our intent is to sketch how a comparison with Vasiliev might be started, rather than to provide such a comparison in detail.)
It is straightforward to check that the generators in \eqref{subalgebra} may be represented as
\beqn\label{rep}
\fr{D}&=& \frac{1}{2}\epsilon^{ij}Y^{d}_iY^{-1}_j\nonumber\\
\fr{P}^a&=& \frac{1}{2}\epsilon^{ij}\left(Y^{a}_iY^{-1}_j+Y^a_iY^{d}_j\right)\\
\fr{M}^{a,b} &=& \frac{1}{2}\epsilon^{ij}Y^{a}_iY^b_j\nonumber
\eeqn
with the Lie-bracket given by the star commutator $[A,B]_{\star} = A\star B-B\star A$. The flat background connection then takes the form
\beq
\cConn^{(0)}(z,x|Y)= -\frac{1}{2}\epsilon^{ij}Y^{d}_iY^{-1}_j\frac{dz}{z}+\frac{1}{2}\epsilon^{ij}\left(Y^{a}_iY^{-1}_j+Y^a_iY^{d}_j\right)\eta_{a\mu}\frac{dx^{\mu}}{z}
\eeq
Similarly, denoting the full connection and pseudoscalar as $\cConn(z,x|Y)$ and $\cScal(z,x|Y)$ respectively, the gauge transformations take the form
\beq\label{RG1}
\delta\cScal = \left[\epsilon, \cScal\right]_{\star},\;\;\;\delta\cConn = \bd\epsilon+\left[\cConn,\epsilon\right]_{\star}
\eeq
where the parameter $\epsilon(z,x|Y)$ is also thought of as a function of the auxiliary variables. We may write the renormalization group equations as
\beqn\label{RG2}
\bd\cScal +\left[\cConn,\cScal\right]_{\star} &=& \Beta^{(\cScal)}_{\star}\nonumber\\
\bd\cConn + \cConn \wedge_{\star} \cConn &=& \Beta^{(\cConn)}_{\star}
\eeqn
where the z-components of the $\star$-beta functions can be read off from equations \eqref{beta1} and \eqref{beta2} after replacing the integral product with the star product, while the transverse components are constrained by Bianchi identities as before. It is here that we see our first real contact with the formalism of Vasiliev\footnote{For readers familiar with the Vasiliev theory, note that Vasiliev equations are usually written in terms of a 0-form $B$ in the \emph{twisted} adjoint representation. If one considers the redefinition $\cScal = B\star \mathcal{K}$ with $\mathcal{K}$ being the Kleinian, then the new 0-form $\cScal$ transforms in the adjoint representation, as opposed to the twisted adjoint.}-- in particular his organization of higher spin gauge fields is seen as a particular representation of the general algebraic structure that arises from consideration of the renormalization group.

\subsection{Ghosts arise upon moving to the Principal bundle}
\begin{figure}[!h]
	\centering
	\includegraphics[width=6cm]{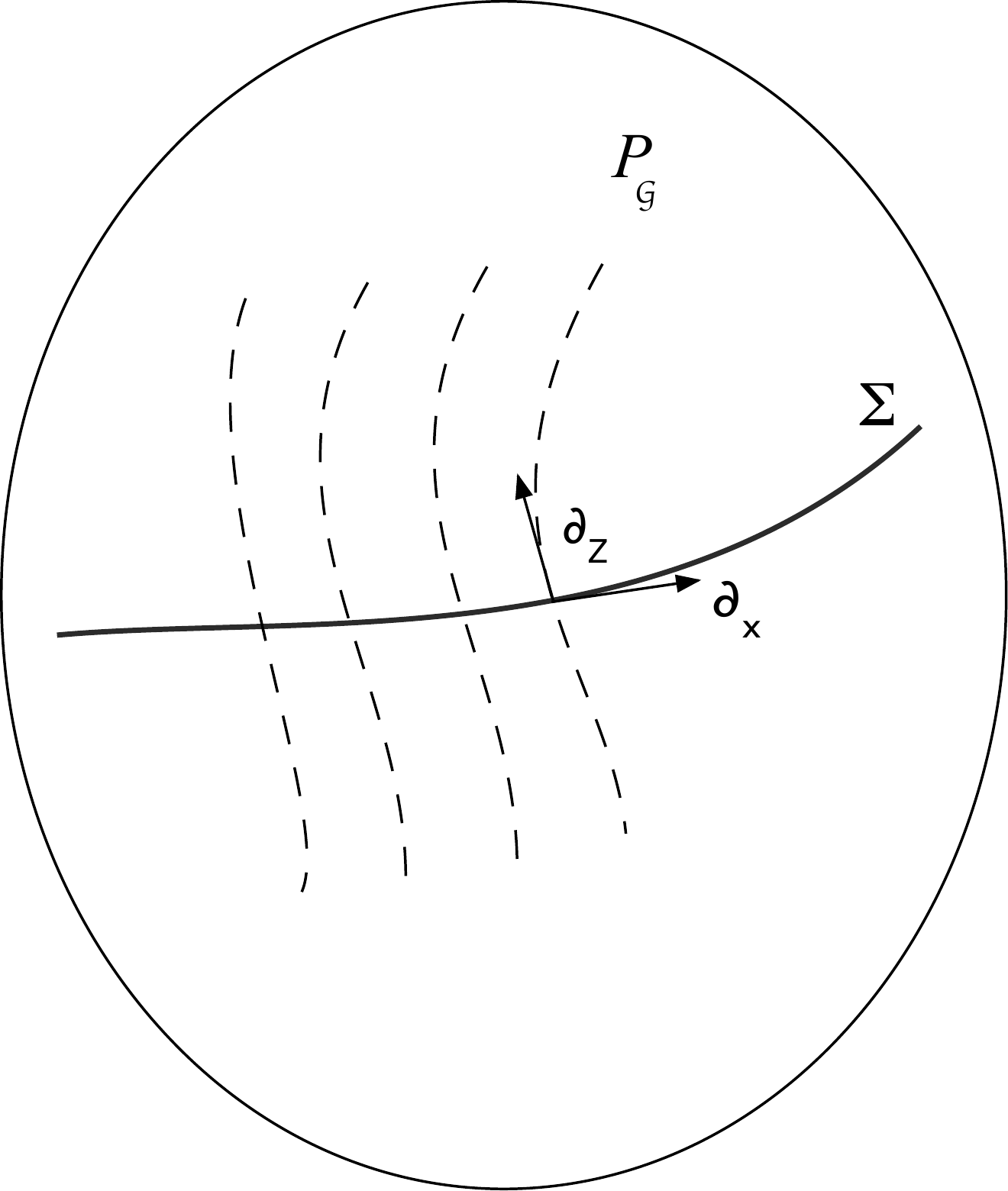}
	\caption{A pictorial representation of the principal bundle structure.}
	\label{fig:gbundle}
\end{figure}

So far we have been thinking of $\cConn$ in terms of a connection on a vector bundle, as is usually the case in most applications in physics. We will now make some observations about the additional structure which we expect to emerge by shifting to the language of principal bundles (see \cite{ThierryMieg:1979kh} for details). Let $\cG\mapsto P_{\cG}\mapsto M_{d+1}$ be a principal bundle over $M_{d+1}$ (with $\cG$ being the structure group), of which $J_{bulk}^{\infty}(E)$ is an associated vector bundle. In particular, we may take $P_{\cG}$ to be the frame bundle $\mathrm{Fr}(J_{bulk}^{\infty}(E))$. Let $Z^{\uM}$ be local coordinates on the (infinite-dimensional) fibers of $P_{\cG}$. Given a local section $\Sigma:M_{d+1}\mapsto P_{\cG}$, we may choose local coordinates\footnote{In this section, the symbol $x$ should be taken to stand for $x^I=(z,x^{\mu})$.} $(x,Z)$ on the total space of $P_{\cG}$ \emph{adapted} to the section, which is to say the section is given by $Z=0$ in these coordinates (see figure \ref{fig:gbundle}). Vector fields on $P_{\cG}$ of the form $V=V^{\uM}\frac{\partial}{\partial Z^{\uM}}$ which point along the fiber directions are referred to as \emph{vertical} vector fields.  

In order to specify what it means to be \emph{horizontal}, we need to define the notion of a connection on the $\cG$-bundle. An \emph{Ehresmann connection} $\PConn$ on $P_{\cG}$ is a $\cG$-equivariant one-form on the total space, valued in the Lie-algebra of $\cG$, and may be written locally on $P_{\cG}$ as 
\beq
\PConn = \PConn_I(x,Z|Y) dx^I + \PConn_{\uM}(x,Z|Y)dZ^{\uM}
\eeq
Note that both $\PConn_I$ and $\PConn_{\uM}$ are valued in the Lie-algebra of $\cG$, which is manifested above by their $Y$ dependence. Having defined the connection, we now refer to vector fields on $P_{\cG}$ in the kernel of $\PConn$ as \emph{horizontal}. In terms of the local coordinate basis of 1-forms $(dx^I,dZ^{\uM})$, we may think of $dx^I$ as being horizontal because they kill all vertical vector fields, while $dZ^{\uM}$ are simply normal to the section $\Sigma$. The pull back of the connection by the section, $\Sigma^{-1}\PConn$, is a qualified connection 1-form on associated vector bundles, and is what is usually called the connection (or gauge field) in the physics literature. It is this piece which may be identified with what we referred to as the connection over $J^{\infty}_{bulk}(E)$ in the previous section
\beq
\cConn(x|Y)=\PConn_I(x,0|Y)dx^I 
\eeq
As was explained in \cite{ThierryMieg:1979kh}, the remaining piece $\PConn_{\uM}(x,0|Y)dZ^{\uM}$ (evaluated on the section) is called the \emph{Faddeev-Popov ghost} in physics, and we suggestively label it as
\beq
S(x|Y)=\PConn_{\uM}(x,0|Y)dZ^{\uM}
\eeq
The fact that $S$ is a 1-form means that it anti-commutes with itself, which is why the ghost is taken to be Grassman. 

The exterior derivative $d$ on the total space $P_{\cG}$ can also be separated with respect to our coordinate system into a horizontal and a vertical piece: $d=\bd_x+d_Z$. The vertical piece $d_Z$ is commonly referred to as the \emph{BRST operator} in physics. The \emph{curvature} 2-form for $\PConn$\footnote{Here $d_ZS$ is to be interpreted appropriately as $d_Z(\PConn_{\uM}dZ^{\uM})|_{Z=0}$.}
\beqn
\cF_{\PConn} &=& d\PConn +\PConn\wedge_{\star} \PConn \nonumber\\
&=& \bd_x\cConn + \cConn\wedge_{\star}\cConn + d_Z\cConn+\bd_xS+\left\{\cConn,S\right\}_{\star}+d_ZS+S\wedge_{\star}S
\eeqn
consequently splits up into a horizontal, a vertical and a mixed term. A fundamental property of the curvature 2-form is that it is purely horizontal (a quick proof for physicists can be found in \cite{ThierryMieg:1979kh}). This implies that the curvature 2-form must not have any $dZ^{\uM}$ legs, which lead us to conclude that
\beq
d_Z\cConn+\bd_xS+\left\{\cConn,S\right\}_{\star}=0
\eeq
\beq
d_ZS+S\wedge_{\star}S=0
\eeq
These relations are referred to as the BRST equations in physics. Of course, the charged 0-form $\cScal$ has its own BRST relation as well, which encodes its tensorial transformation property under gauge transformations
\beq
d_Z\cScal + \left[S,\cScal\right]_{\star} = 0
\eeq
At this point, putting all of the above BRST equations together with the renormalization group equations \eqref{RG1} and \eqref{RG2}, we obtain the full set of equations satisfied by the various pieces of our Ehresmann connection
\beqn\label{Gbundleequations}
& &\bd_x\cConn+\cConn\wedge_{\star}\cConn = \Beta^{(\cConn)}_{\star}\nonumber\\
& &\bd_x\cScal+\left[\cConn, \cScal\right]_{\star} = \Beta^{(\cScal)}_{\star}\nonumber\\
& &d_Z\cConn+\bd_xS+\left\{\cConn,S\right\}_{\star}=0\\
& &d_Z\cScal + \left[S,\cScal\right]_{\star} = 0\nonumber\\
& &d_ZS+S\wedge_{\star}S=0\nonumber
\eeqn
These equations bear remarkable resemblance with the equations of motion in Vasiliev's higher spin theory, which have been briefly reviewed for completeness in Appendix \ref{sec: vasiliev}. 
Note however, that there are also significant differences: 

(i) Firstly, in our construction, $Z^{\uM}$ are coordinates on the infinite dimensional fibers of $P_{\cG}$. To make contact with Vasiliev, we can introduce a parameterization of these fiber coordinates
\beq
Z^\alpha=\sum\left( z^\alpha_{A_1B_1...}\epsilon^{i_1j_1}Z^{A_1}_{i_1}\star Z^{B_1}_{j_1}\star...\right)
\eeq
That is, by introducing auxiliary $sp(2)\times O(2,d)$ variables $Z^A_i$, the $Z^\alpha$ can be written as arbitrary $Sp(2)$-invariant $\star$-polynomials. We can then recast
\beq\label{LiftS}
S(x|Y)=\PConn_{\uM}(x,0|Y)dZ^{\uM}
=\PConn_A^i(x|Y,Z)dZ^A_i
\eeq

(ii) Secondly, equations \eqref{Gbundleequations} have been written along the $Z^\alpha=0$ section, and (\ref{LiftS}) represents some sort of lift to non-zero $Z^A_i$. While one is eventually supposed to project the non-linear Vasiliev equations to $Z^A_i=0$ to get the physical variables, such a projection is not straightforward in the Vasiliev theory, and is typically carried out order by order in perturbation theory, thus making a direct comparison non-trivial.

(iii) Finally, in Vasiliev's equations without the projection to $Z^A_i=0$, the curvature is along vertical (i.e. $dZ^A_i\wedge dZ_A^i$) directions, as opposed to our situation, where the horizontal components of curvature are non-trivial. 

It is natural to ask if there is some sort of redefinition of our variables that would render our equations in Vasiliev's form. Such a redefinition was implicit in the construction of Ref. \cite{Douglas:2010rc}, though it is not clear to us if such a redefinition is natural. From our point of view, it seems compelling to think of the RG $\beta$ functions as the (horizontal) curvature, while the equations for $S$ are interpreted as the analogue of BRST equations. Holographic RG certainly presents us with a notion of a higher spin theory; it is perhaps not obvious that it must agree in all details with Vasiliev's construction, even though the similarities are immense. However, it is our belief that the differences pointed out above conspire to hide the equivalence of our renormalization group equations with the non-linear Vasiliev equations. A better understanding of this equivalence by constructing an explicit map between the two sets of equations will be left to future work. But if the conjectured equivalence is indeed true, then it would shed new light on the auxiliary 1-form $S$ in the Vasiliev system (which has always appeared mysterious, to us anyway), namely, that it is the Faddeev-Popov ghost corresponding to the higher-spin gauge symmetry.

Let us end this section with a comment on the usual Faddeev-Popov formalism in physics. Conventionally, one quantizes a classical gauge theory, thought of as a theory of the horizontal components of the connection, by integrating over equivalence classes of such connections. This is usually described as dividing the path integral measure by the volume of the gauge group. The Faddeev-Popov ghosts enter upon gauge-fixing. The resulting free path integral may be interpreted to mean that the quantum theory should be considered as an integration over connections on the principal bundle. Apparently, we are in a somewhat analogous situation here. The field theory has sources corresponding to the horizontal pieces of the connection, while the ghosts are absent. As we have seen, it is certainly natural to introduce the ghosts to complete the geometrical structure, but whether it is inevitable, is far from clear. Perhaps further thought along the lines of Batalin-Fradkin-Vilkovisky theory would be fruitful.\footnote{We thank D. Minic for pointing out the importance of BFV in string field theory \cite{Zwiebach:1993cs}, which may be related.}

\section{Discussion}

\subsection{The Bosonic Theory}\label{sec:bosonic}

We can write the bosonic $O(N)$ theory in a similar way to the Majorana model, although it is not nearly as simple. The trick is to recognize that the kinetic term can be written in terms of the $P_{F;\mu}$ that we introduced in the fermionic theory above. This should be expected because of its geometric significance. Indeed, we choose to write the action coupled to arbitrary $O(N)$ singlet operators in the matrix form
\beq\label{Bosaction}
S^{reg.}_{bos}[\phi,A,W]=\int_{x,y}\Big[
\frac12\int_z\phi^m (x)D_{F;\mu}(x,z)D_{F;}{}^{\mu}(z,y)\phi^m(y)-\frac12\phi^m(x)B(x,y)\phi^m(y)\Big]
\eeq
where
\beq
D_{F;\mu}(x,y)=P_{F;\mu}(x,y)+W_\mu(x,y)
\eeq
This is equivalent up to some redefinitions to the paramaterization employed in \cite{Douglas:2010rc}, although it is clearly more suited to the geometric interpretation. 
Thus we see that without loss of generality, the (singlet sector of the) bosonic theory can be thought of as consisting of sources $W^\mu$ and $B$. We note that $B$ is parity even here (recall that the scalar source $A$ in the Majorana theory was parity odd), a fact familiar from the structure of the Vasiliev higher spin theories.

The construction developed for the Majorana theory will go through in this case in a very similar fashion, with only the $\beta$ functions (and thus the three point functions) being modified appropriately. The details of this construction will not be given here for brevity. 

We note though that in the bosonic case, the jet bundle construction will go through in arbitrary dimension. This is perhaps related to the fact that there is a vectorial construction of the Vasiliev A-model in any dimension. In the fermionic case, the B-model is known only in four bulk dimensions in the twistorial construction. The RG analysis can of course be carried out in arbitrary dimensions, suggesting that corresponding higher spin theories do exist. However, we note that such theories would be more complicated than the Vasiliev theories, as we must include sources for operators of the form $\tpsi \gamma^{ab...}\psi$, corresponding to higher tensor fields in the bulk. The jet bundle construction suggests that these cannot be absorbed into the connection.

\subsection{Interacting theories}\label{Sec:interactions}

It is a familiar idea that RG fixed points correspond to zeroes of the RG $\beta$ functions. We have remarked previously that the connection $W^{(0)}$ that can be identified with $AdS_{d+1}$ is an exact solution of the full set of ERG equations, corresponding to the point $(A,\widehat{W}_\mu)=(0,0)$. This indeed corresponds to the only zero of $\beta^{(A)}$ and $\beta^{(W)}_\mu$. Other fixed points of RG might arise once field theory interactions are turned on.

Indeed, given the analysis based on the free fixed point, it is natural to ask what modifications might be expected once interactions are introduced. A natural way to address this would be to introduce sources for non-quadratic operators. However in doing so, we would immediately lose much of the geometric structure that we have described. 

Another way to introduce a large class of interactions is to implement the reverse of the Hubbard-Stratanovich idea. Namely, given $Z[M,z,W_\mu,A]$, we construct the partition function of interacting theories by integrating over the source with suitable weight. 
To proceed, we choose a gauge in which $A$ is diagonal
\beq
A(x,y)=\sigma(x)\delta(x-y)
\eeq
and write (one could in addition introduce a source for $\sigma$)
\beq\label{InteractingTransform}
Z^{int}[M,z,W_\mu]=\int [d\sigma(x)]e^{i\frac{N}{2g}\int d^dx\ \sigma^2}Z[M,z,W_\mu,\sigma]
\eeq
This transform corresponds to the particular case in which we implement a `double-trace' deformation. It is precisely at this point at which large $N$ matters. In particular, at large $N$, the integral can be done by saddle point approximation. The corresponding gap equation is 
\beq
\frac{N}{g}\sigma+\Pi=0.
\eeq
and one expects to obtain the familiar result that the partition function of the interacting fixed point is obtained essentially as a Legendre transform with $A$ and $\Pi_A$ swapping their roles.

This result seems consistent with the persistence of higher spin symmetry at $N=\infty$ for interacting fixed points. It is far however from an explicit derivation of the RG equations for the interacting fixed point, although it does suggest that at zeroth order in $1/N$, we should just interchange $\Pi_A$ and $A$. At finite $N$, there is every reason to believe that the standard lore would emerge, namely that the higher spin symmetry will be Higgsed in the presence of interactions in the field theory, presumably through an instability towards the condensation of $A$. It would of course be of great interest to find an explicit 'attractor mechanism' in which a purely gravitational theory (presumably in the case where translational invariance in the transverse space is preserved) emerges in the infrared.
Certainly one might expect that the inclusion of field theory interactions might lead to a replacement of our Hamiltonian by a version non-linear in momenta, perhaps along the lines of the construction of Sung-Sik Lee \cite{Lee:2013dln}.

 \section{Acknowledgments}
 
 We enthusiastically thank D. Minic, L. Pando Zayas and D. Vaman for extensive discussions that led to this work, and to the Aspen Center for Physics for providing a stimulating working environment. We are also grateful to Sung-Sik Lee for discussions,  to Pierre Albin for discussions about jet bundles, and again to D. Minic for comments and suggestions on a draft of this paper. We thank E. Sezgin and E. Skvortsov  for comments on the first version of the paper. Research supported by the US Department of Energy under contract DE-FG02-13ER42001. 
 

\appendix\numberwithin{equation}{section}
\section{Renormalization group: details}\label{RGappendix}

In this appendix, we present the details of RG equations and Callan-Symanzik equations. We will derive these equations by going through the 2-step RG transformation explained in Section \ref{section:RG}. In doing so, we will find it convenient to split the action as
\beqn
	S_{reg.}^{Maj} &=& S_{0}+ S_{int}+\pot, \nonumber\\
	S_{0} &=& \int_{x,y} \tpsi^a(x) i\slashed D^{(0)}(M;x,y)\psi^a(y),
	                  \quad \left\{ \begin{array}{ll}D^{(0)}_{\mu}(M;x,y) = P_{F;\mu}(M;x,y) - i\Conn_{\mu}^{(0)}(x,y)\\
	                                                             P_{F;\mu}(M;x,y) = K_F^{-1}(-\Box_x/M^2)\partial^x_{\mu}\delta(x-y)\end{array} \right. \nonumber\\
	S_{int} &=& \int_{x,y}\tpsi^{a}(x)\left[\Scal(x,y) + \slashed\Cont(x,y)\right]\psi^a(y) \label{Sint}, \nonumber\\
	\pot &=& \int_{x,y}\potfunc(x,y) \equiv \int_{x,y} \potfunc_0\,\delta(x-y)
\eeqn
where recall that ${\tpsi}^a_\beta\equiv \psi^{a;\alpha}\epsilon_{\alpha\beta}$ is not an independent field.

\subsection*{Exact RG equations}
\textbf{Step 1}: We first begin by lowering the cut-off from $M$ to $\lambda M$ for $\lambda <1$. From the Wilsonian point of view, this essentially amounts to integrating out a shell of fast modes. The way to carry out this integration within the Polchinski formalism, is to demand
\beq
	Z[M,z, \Scal, \Conn_{\mu}, \potfunc] = Z[\lambda M,z, \widetilde\Scal, \widetilde\Conn_{\mu}, \widetilde\potfunc] \label{RGfinite}
\eeq
What is being said here, is that we're adjusting the values of the sources (denoted by tilde) in order to keep the path integral unchanged. Infinitesimally, Taking $\lambda=1-\epsilon$ in \eqref{RGfinite} gives
\beq
	\label{Polchinski1}
	0 = \delta_\epsilon Z 
	   = \delta_\epsilon \left(Z_0^{-1} \!\int[d\psi]e^{iS} \right)
	   =  - Z_0^{-1} \left( \delta_\epsilon \int[d\psi]e^{iS_0} \right) Z_0^{-1} \, \int[d\psi]e^{iS} + Z_0^{-1} \, \left( \delta_\epsilon \int[d\psi]e^{iS} \right)
\eeq
with
\beqn
	\delta_\epsilon \int[d\psi]e^{iS_0}
	&=& \int[d\psi] \rgla e^{iS_0} \nonumber\\
	\delta_\epsilon \int[d\psi]e^{iS}
	&=& \int[d\psi] \left(e^{i(S_{int}+\pot)}\rgla e^{iS_0}
	                              + e^{iS_0}\mathrm{Tr}\!\!\left\{\delta_{\epsilon}\Scal\bl\frac{\delta}{\delta\Scal} - \delta_{\epsilon}\Cont_{\mu}\bl\frac{\delta}{\delta\Cont_{\mu}}
	                                                                + \delta_\epsilon\potfunc\bl\frac{\delta}{\delta\potfunc}\right\} e^{i(S_{int}+\pot)}\right) \nonumber\\
\eeqn
where $\mathrm{Tr} f(u,v)\equiv\int_{u,v}\delta(u-v)f(u,v)$ is the functional trace.  It is convenient to define
\beq
\slashed\alp(x,y) \equiv \gamma^{\mu}\alp_{\mu}(x,y) = \rgla \left(i\slashed D^{(0)}\right)^{-1} \!(x,y)
\eeq
Given our choice of \(S_0\), we get
\beq
	\label{maj1}
	\rgla e^{iS_0} 
	= -i \int_{x,y} \left(i\slashed{D}^{(0)}\bl\psi\right)^{\alpha}(x)\epsilon_{\alpha\beta}
	      {\left(\slashed\alp\right)^{\beta}}_{\gamma}(x,y) \left(i\slashed{D}^{(0)}\bl\psi\right)^{\gamma}(y) \ e^{iS_0}
\eeq
where we have supressed the $O(N)$ vector indices, and explicitly shown some of the spinor indices.
Using 
\beq
\epsilon^{\alpha\beta}\frac{\delta S_0}{\delta \psi^{\beta}(x)}=2\left(i\slashed{D}^{(0)}\bl \psi\right)^{\alpha}(x)
\eeq
we may re-write \eqref{maj1} as
\beq
	\rgla e^{iS_0} = -\frac{i}{4}\int_{x,y}{(\slashed\alp)^{\alpha}}_{\gamma}(x,y)\epsilon^{\gamma\beta}\left(\frac{\delta^2}{\delta \psi^\alpha(x)\delta \psi^\beta(y)}e^{iS_0}-i\frac{\delta^2S_0}{\delta \psi^\alpha(x)\delta \psi^\beta(y)}e^{iS_0}\right)
\eeq
Plugging this back into \eqref{Polchinski1}, canceling terms and integrating by parts, we arrive at
\beq
	\label{maj2}
	\mathrm{Tr}\!\!\left\{\delta_{\epsilon}\Scal\bl\frac{\delta}{\delta\Scal} - \delta_\epsilon\Cont_{\mu}\bl\frac{\delta}{\delta\Cont_{\mu}}
	                  + \delta_\epsilon\potfunc\bl\frac{\delta}{\delta\potfunc}\right\} e^{i(S_{int}+U)}
	= \frac{i}{4}\int_{x,y}{(\slashed\alp)^\alpha}_{\gamma}(x,y)\epsilon^{\gamma\beta}
	    \frac{\delta^2}{\delta\psi^{\alpha}(x)\delta\psi^{\beta}(y)}e^{i(S_{int}+U)}
\eeq
Using the explicit form of $S_{int}$ and $U$ from equation \eqref{Sint} we then find
\beq
	\left[\delta_{\epsilon}\Scal\;\delta^{\alpha}_{\beta}+\delta_{\epsilon}\Cont_{\mu}\;{(\gamma^{\mu})^{\alpha}}_{\beta}\right]
	= \left[ \Scal\;{\delta^{\alpha}}_{\gamma}+ \Cont_{\mu}\;{(\gamma^{\mu})^{\alpha}}_{\gamma}\right]\bl {(\slashed\alp)^{\gamma}}_{\delta}
	      \bl\left[ \Scal\;{\delta^{\delta}}_{\beta}+ \Cont_{\nu}\;{(\gamma^{\nu})^{\delta}}_{\beta}\right]
\eeq
\beq
	i\delta_\epsilon U = \frac{N}{4}\int_{x,y} {(\slashed\alp)^\alpha}_\beta(x,y){(A+\slashed\Cont)^\beta}_\alpha(x,y)
\eeq
Now restricting our attention to 2+1 dimensions, we evaluate the various gamma matrix products on the right hand sides of the above two equations. Then comparing the spinor matrix structure on both sides, we obtain
\beq
	\delta_\epsilon \Scal \equiv \beta_\Scal
	=\Scal\bl \alp^\mu\bl \Cont_\mu + \Cont_\mu \bl \alp^\mu \bl \Scal + \epsilon^{\mu\nu\lambda}\Cont_\mu \bl \alp_\nu \bl \Cont_\lambda
\eeq
\beq
	\delta_\epsilon \Cont_\mu \equiv \beta_{\Conn;\mu}
	=\Scal \bl \alp_\mu \bl \Scal + \epsilon_{\mu\nu\lambda} \left( \Scal \bl \alp^\nu \bl \Cont^\lambda + \Cont^\nu \bl \alp^\lambda \bl \Scal \right)
	   + \Cont_\nu \bl \alp^\nu \bl \Cont_\mu - \Cont_\nu \bl \alp_\mu \bl \Cont^\nu + \Cont_\mu \bl \alp_\nu \bl \Cont^\nu
\eeq
\beq
	\delta_\epsilon \pot \equiv \beta_{\pot}= -i \frac{N}{2} \, \mathrm{Tr}\!\left\{ \alp_\mu \bl \Cont^\mu \right\}
\eeq

\textbf{Step 2}: Next, we perform a $CO(L_2)$ scale transformation, accompanied by an arbitrary spatial translation, $\cL = 1 + \varepsilon zW_z + \varepsilon\xi^\mu W_\mu$, such that the partition function comes back to the original cut-off $M$, but the conformal factor of the background metric changes as $z\mapsto \lambda^{-1}z$. We then label the final sources as $\Scal(\lambda^{-1}z;x+\varepsilon\xi,y+\varepsilon\xi)$, $\Conn_{\mu}(\lambda^{-1}z;x+\varepsilon\xi,y+\varepsilon\xi)$ and $\pot(\lambda^{-1}z)$, which are given by
\beq
	\Scal(\lambda^{-1}z;x+\varepsilon\xi,y+\varepsilon\xi)
	  = \Scal(z;x,y) - \varepsilon z \left[ \Conn_z , \Scal \right]_\bl - \varepsilon\xi^\mu \left[ W_\mu , A \right]_\bl + \varepsilon \beta_\Scal
	      + O(\varepsilon^2)
\eeq
\beq
	\Conn_\mu(\lambda^{-1}z;x+\varepsilon\xi,y+\varepsilon\xi)
	  = \Conn_\mu (z) + \varepsilon z \left[ P_{F;\mu} + \Conn_\mu , \Conn_z \right]_\bl + \varepsilon\xi^\nu \left[ P_{F;\mu}
	      + \Conn_\mu , \Conn_\nu \right]_\bl + \varepsilon \beta_{\Conn;\mu} + O(\varepsilon^2)
\eeq
\beq
	\pot(\lambda^{-1}z)
	  = \pot(z) +  \varepsilon \beta_{\pot} - i \varepsilon \frac{N}{2} \mathrm{Tr}\;\Big\{\alp_z \bl \Conn^z\Big\}
\eeq
where we have introduced the notation $\alp_z$ to denote a (possible) $CO(L_2)$ anomaly.  In particular, $\alp_z$ should be thought of as the anomaly for a single Majorana fermion, hence the scaling of the full anomaly with $N$. Note that given the structure of $\beta_{\pot}$, it seems as if $\alp_z$ naturally combines with $\alp_{\mu}$ into $\alp_I=(\alp_z,\alp_\mu)$.  Finally, expanding out the left hand sides of the above relations and taking $\varepsilon$ to zero, we arrive at the ERG equations \eqref{MajRG1} and \eqref{MajRG2}.

\subsection*{Callan-Symanzik equation}

We are also interested in the Callan-Symanzik equations for quadratic operators like \(\hat{\ScalM}(x,y)=\frac{1}{2}\psi^{\alpha}(x)\epsilon_{\alpha\beta}\psi^{\beta}(y)\) and \(\hat{\ConnM}^{\mu}(x,y)=\frac{1}{2}\psi^{\alpha}(x)\epsilon_{\alpha\beta}{(\gamma^{\mu})^{\beta}}_{\delta}\psi^{\delta}(y)\). For a generic operator \(\mathcal{O}\), one can straightforwardly check from an argument similar to the one described above, that
\beq
	\rgla \langle \cO\rangle
	   = \frac{1}{4} \int_{u,v} {\alp^\gamma}_\delta(u,v) \epsilon^{\delta\eta}
	       \left\langle -\frac{\delta S_{int}}{\delta \psi^\gamma(u)}\frac{\delta \cO}{\delta\psi^\eta(v)}
	       - \frac{\delta \cO}{\delta\psi^\gamma(u)} \frac{\delta S_{int}}{\delta \psi^\eta(v)}
	       + i \frac{\delta^2\cO}{\delta\psi^\gamma(u)\delta\psi^\eta(v)} \right\rangle
\eeq
For the case of quadratic interactions, as before we have
\beq
	\frac{\delta S_{int}}{\delta\psi^{\gamma}(u)}
	   = 2 \int_z \epsilon_{\gamma\beta} \left[ \Scal(u,z){\delta^{\beta}}_{\delta} + \Cont_{\mu}(u,z) {(\gamma^\mu)^\beta}_\delta \right]\psi^{\delta}(z)
\eeq
Let us also consider the general quadratic operator \(\mathcal{O}_M = \psi^{\alpha}(x)\epsilon_{\alpha\beta}{M^{\beta}}_{\delta}\psi^{\delta}(y)\). We have
\beq
	\frac{\delta\mathcal{O}_M}{\delta\psi^{a;\gamma}(u)}
	   = \delta^{(d)}(x-u) \epsilon_{\gamma\beta} {M^\beta}_\delta \psi^{a;\delta}(y) - \psi^{a;\alpha}(x)\epsilon_{\alpha\beta} {M^\beta}_\gamma \delta^{(d)}(y-u),
\eeq
\beq
	\frac{\delta^2\mathcal{O}_M}{\delta\psi^{a;\gamma}(u)\delta\psi^{a;\eta}(v)}
	   = N \left( \delta^{(d)}(x-u) \delta^{(d)}(y-v) \, \epsilon_{\eta\beta} {M^\beta}_\gamma
	                  - \delta^{(d)}(x-v) \delta^{(d)}(y-u) \, \epsilon_{\gamma\beta} {M^\beta}_\eta \right),
\eeq
where the $N$ appears from tracing over $O(N)$ indices.  Thus, after step one of RG we get
\begin{align}
	\delta_{\varepsilon} \langle\mathcal{O}_M\rangle
	   &= - i \frac{N}{2} {\alp^\beta}_\gamma(x,y) {M^\gamma}_\beta \nonumber\\
	         &\quad
	         - \int_{u,v,z} \left\langle  \psi^{\kappa}(z) \epsilon_{\kappa\rho} \left[ \Scal(z,v) \delta^{\rho}_{\eta} + \Cont_{\mu}(z,v){(\gamma^\mu)^\rho}_\eta \right]
	                               {\alp^\eta}_{\delta}(v,x) \epsilon^{\delta\gamma} \epsilon_{\gamma\beta} {M^\beta}_\tau \psi^\tau(y)  \right\rangle \nonumber\\
	         &\quad
	         - \int_{u,v,z} \left\langle  \psi^\alpha(x) \epsilon_{\alpha\beta} {M^\beta}_\gamma {\alp^\gamma}_\delta(y,v) \epsilon^{\delta\eta} \epsilon_{\eta\rho}
	                               \left[\Scal(v,z)\delta^\rho_\kappa + \Cont_\mu(v,z) {(\gamma^\mu)^\rho}_\kappa \right] \psi^\kappa(z)  \right\rangle.
\end{align}
We may now write down separate equations for \(M\) either 1 or \(\gamma^{\mu}\). Since both the operators transform tensorially under $CO(L_2)$, after step 2 we get:
\begin{align}
	 \ScalM(z+\varepsilon z; x+\varepsilon\xi,y+\varepsilon\xi)
	&=  \ScalM(z; x,y)+\left[\ScalM, \varepsilon z\Conn_z+\varepsilon\xi^{\mu}\Conn_{\mu}\right]\nonumber\\
	&+\varepsilon\left( \alp_\nu \bl \Scal \bl \ConnM^\nu - \ConnM^\nu \bl \Scal \bl \alp_\nu \right)
	      - \varepsilon\left( \alp^\mu \bl \Cont_\mu \bl \ScalM + \ScalM \bl \Cont_\mu \bl \alp^\mu \right) \nonumber\\
	      &\quad
	      + \varepsilon\epsilon^{\mu\nu\lambda} \left( \alp_\mu \bl \Cont_\nu \bl \ConnM_\lambda + \ConnM_\mu \bl \Cont_\nu \bl \alp_\lambda \right)+O(\varepsilon^2)
\end{align}
\begin{align}
	\ConnM^\mu(z+\varepsilon z; x+\varepsilon\xi,y+\varepsilon\xi)
	 &=\ConnM^{\mu}(z; x,y)- i \varepsilon N \alp^\mu + \left[\ConnM^{\mu}, \varepsilon z\Conn_z+\varepsilon\xi^{\mu}\Conn_{\mu}\right]\nonumber\\
	     & + \varepsilon\left( \alp^\mu \bl \Scal \bl \ScalM + \ScalM \bl \Scal \bl \alp^\mu \right)
	                 + \varepsilon\epsilon^{\mu\nu\sigma} \left( \alp_{\nu} \bl \Scal \bl \ConnM_{\sigma}
	                 + \ConnM_\nu \bl \Scal \bl \alp_\sigma \right) \nonumber\\
	      &\quad
	      - \varepsilon\left( \alp^\nu \bl \Cont_\nu \bl \ConnM^\mu + \ConnM^\mu \bl \Cont_\nu \bl \alp^\nu \right)
	      - \varepsilon\left( \alp_\nu \bl \Cont^\mu \bl \ConnM^\nu + \ConnM^\nu \Cont^\mu \alp_\nu \right) \nonumber\\
	      &\quad
	      - \varepsilon\left( \alp^\mu \bl \Cont_\nu\bl \ConnM^\nu + \ConnM^\nu \bl \Cont_\nu \bl \alp^\mu \right)
	      + \varepsilon\epsilon^{\mu\nu\lambda} \left( \alp_\nu \bl \Cont_\lambda \bl \ScalM+ \ScalM \bl \Cont_\nu \bl \alp_\lambda \right)
\end{align}
The Callan-Symanzik equations can be written in a more compact form by making the definitions  
\beq
\gamma(x,y;u,v) = \delta(x-u)\alp^{\mu}\bl \Cont_{\mu}(y,v)+\Cont_{\mu}\bl\alp^{\mu}(u,x)\delta(v-y)
\eeq
\beqn
\gamma^{\mu}(x,y;u,v) &=& 
\delta(u-x)\alp^{\mu}\bl A(y,v)
+A\bl \alp^{\mu}(u,x)\delta(v-y)\nonumber\\
&+&
\epsilon^{\mu\nu\lambda}\left(\delta(x-u)\alp_{\nu}\bl\Cont_{\lambda}(y,v)
+\Cont_{\nu}\bl \alp_{\lambda}(u,x)\delta(v-y)\right)
\eeqn
\beqn
\gamma_{\mu\nu}(x,y;u,v) &=& 
\epsilon_{\mu\lambda\nu}\delta(x-u)\alp^{\lambda}\bl A(y,v)
+\epsilon_{\nu\lambda\mu}A\bl \alp^{\lambda}(u,x)\delta(v-y)
\nonumber\\
&+& 
\delta(x-u)\alp_{\mu}\bl\Cont_{\nu}(y,v)
+\Cont_{\nu}\bl\alp_{\mu}(u,x)\delta(y-v)
\nonumber\\
&-& 
\delta(x-u)\alp_{\nu}\bl\Cont_{\mu}(y,v)
-\Cont_{\mu}\bl\alp_{\nu}(u,x)\delta(y-v)
\nonumber\\
&+&
\delta(x-u)\alp_{\lambda}\bl\Cont^{\lambda}(y,v)\eta_{\mu\nu}
+\Cont_{\lambda}\bl\alp^{\lambda}(u,x)\delta(y-v)\eta_{\mu\nu}
\eeqn
Having done so, comparing the $\varepsilon$  terms on both sides, we obtain the Callan Symanzik equations \eqref{MajCS1} and \eqref{MajCS2}.

\section{Vasiliev Higher spin gravity}\label{sec: vasiliev}
\newcommand{\ga}{\alpha}
\newcommand{\dga}{\dot\alpha}
\newcommand{\gb}{\beta}
\newcommand{\dgb}{\dot\beta}
\newcommand{\gep}{\epsilon}
\newcommand{\gK}{\mathcal{K}}
In this section, we will present a short review of the non-linear Vasiliev higher spin equations in general dimension $d+1$ in terms of vector oscillators.\footnote{We note that the case $d=3$ is special, in that the Vasiliev equations can be formulated in terms of twistor  variables, and admit the two versions referred to as A type and B type. In particular, it is not known how to construct the B type theory in terms of vector oscillators. We do not wish to confuse the reader on this point (it is the $d=3$ B model that is directly addressed in this paper) --- we merely provide this Appendix as an introduction to some of the language that we used in the body of the paper.} Of course, this is not meant to be pedagogical by any means, as the details are not relevant to our discussion in this paper -- our aim here is to merely present the Vasiliev equations so as to facilitate comparison with our RG equations. For more details on the Vasiliev theory, we refer the reader to Refs. \cite{Vasiliev:1995dn, Vasiliev:2012vf, Vasiliev:1999ba, Bekaert:2005vh, Giombi:2012ms}.

Let \(\{Y^A_{i}\}\) and \(\{Z^A_{j}\}\) be \(Sp(2)\times O(2,d)\) variables, where upper-case latin indices \(A,B\cdots\) stand for \(O(2,d)\)  vector indices, while \(i,j,\cdots\) stand for \(Sp(2)\) indices. The \(Sp(2)\) invariant product is defined by \(Y^{A\;i}Y^B_i \equiv \epsilon^{ij}Y^A_iY^B_j\). 
We define the \emph{star-product} between two functions \(f(Y,Z)\) and \(g(Y,Z)\) as
\beq
f(Y,Z)\star g(Y,Z) = N^{2D}\int d^{2D}Ud^{2D}V\; e^{-2U^A_iV_A^i}f(Y+U,Z+U)g(Y+V,Z-V)
\eeq
where $D=d+2$ and \(N^{2D}\) is an appropriate normalization constant chosen such that \(f\star 1=f\). It is easy to check that this implies the relations
\beqn
Y^A_i\star Y^B_j&=&Y^A_iY^B_j+\frac{1}{2}\eta^{AB}\epsilon_{ij},\;\;Z^A_i\star Z^B_j=Z^A_iZ^B_j-\frac{1}{2}\eta^{AB}\epsilon_{ij}\nonumber\\
Y^A_i\star Z^B_j &=& Y^A_iZ^B_j-\frac{1}{2}\eta^{AB}\epsilon_{ij},\;\;Z^A_i\star Y^B_j = Z^A_iY^B_j+\frac{1}{2}\eta^{AB}\epsilon_{ij}
\eeqn
We introduce the function \(\gK(t) = e^{-2tz^iy_i}\), where \(y_i = Y^{-1}_i\) and \(z_i = Z^{-1}_i\). For \(t=1\) this is called the \emph{Kleinian}, and will be denoted by \(\gK\). It has the important property that
\beq
\gK\star \gK=1,\;\;\gK \star f(Y,Z)\star \gK = \tilde{f}(Y,Z)
\eeq
where \(\tilde{f}(Y,Z) = f(Y^A-2Y^{-1}\delta^A_{-1}, Z^{A}-2Z^{-1}\delta^A_{-1})\). \newcommand{\hA}{\hat{\mathcal{A}}}
\newcommand{\hF}{\hat{\mathcal{F}}}
\newcommand{\wT}{\widetilde}
\newcommand{\wH}{\widehat}
\newcommand{\cS}{\mathcal{S}}

The Vasiliev system is described by two one forms \(\cConn(x|Y,Z)=\cConn_{I}(x|Y,Z)dx^I\) and \(\cS(x|Y,Z)=S_A^i(x|Y,Z)dZ^{A}_i\), and a zero-form \(B(x|Y,Z)\). The Vasiliev equations are given by 
\beqn\label{VasilievEOM}
& &d_x\cConn+\cConn\star\cConn=0\nonumber\\
& &d_xB+ \cConn\star B-B\star\widetilde{\cConn}=0\nonumber\\
& &d_Z\cConn+d_x\cS+\cConn\star \cS+ \cS\star\cConn = 0\\
& &d_ZB+\cS\star B-B\star\wT \cS=0\nonumber\\
& &d_Z\cS+ \cS\star \cS = \frac{2}{3}dZ^{-1}_idZ_{-1}^{i}\;B\star\gK\nonumber
\eeqn 
In addition, one must impose the appropriate $Sp(2)$ invariance constraints on the above fields, in order for them to describe physical higher spin fields. Note that $B$ transforms in the \emph{twisted} adjoint representation, and in particular the covariant derivatives for $B$ feature the twisted commutators $(\cConn\star B-B\star\widetilde{\cConn})$ and $(\cS\star B-B\star\widetilde{\cS})$. By redefining the 0-form as
\beq
\cScal = B\star \gK
\eeq
the new 0-form $\cScal$ transforms in the adjoint representation, and the twisting can be partially removed from the Vasiliev equations.

\providecommand{\href}[2]{#2}\begingroup\raggedright\endgroup

\end{document}